\documentclass{aa}  

\usepackage{graphicx}
\usepackage{txfonts}
\usepackage{graphicx}	% Including figure files
\usepackage{amsmath,thmtools}	% Advanced maths commands
\usepackage{amssymb}	% Extra maths symbols
\usepackage{xspace} 
\usepackage{epstopdf}
\usepackage{hyperref}
\usepackage{enumitem}
\usepackage{xcolor}
\usepackage{soul}
\sethlcolor{red}
\usepackage{comment}

\newcommand{\msun}{\hbox{$\hbox{\rm M}_{\odot}$}\xspace}
\newcommand{\lsun}{\hbox{${\rm L}_{\odot}$}\xspace}

\newcommand{\kms}{\rm km\,s^{-1}}

\newcommand{\Mpc}{{\rm Mpc}\xspace}

\newcommand{\angstrom}{\mbox{\normalfont\AA}}
\newcommand{\lya}{${\rm Ly\alpha}$\xspace}

\begin{document} 

   \title{Circumgalactic Ly$\boldsymbol\alpha$ emission around submillimeter-bright galaxies with different quasar contributions}
   \author{Vale Gonz\'alez Lobos\inst{1}
          \and
          Fabrizio Arrigoni Battaia\inst{1}
          \and 
          Seok-Jun Chang\inst{1}
          \and 
          Max Gronke\inst{1}
          \and 
          Guinevere Kauffmann\inst{1}
          \and 
          Chian-Chou Chen\inst{2}
          \and 
          Hai Fu\inst{3}
          \and 
          Aura Obreja\inst{4}
          \and
          Emanuele P. Farina\inst{5}
          }
          \titlerunning{Large-scale Ly$\alpha$ around $z\sim4$ submillimeter-bright galaxies}

   \institute{Max-Planck-Institut f\"ur Astrophysik, Karl-Schwarzschild-Str 1, 85748 Garching bei M\"unchen, Germany\\
              \email{valegl@mpa-garching.mpg.de}
         \and
             Academia Sinica Institute of Astronomy and Astrophysics, No. 1, Sec. 4, Roosevelt Rd., Taipei 10617, Taiwan
        \and
            Department of Physics \& Astronomy, University of Iowa, Iowa City, IA 52242, USA
        \and
            Universit\"ats-Sternwarte M\"unchen, Scheinerstraße 1, D-81679 M\"unchen, Germany
        \and
            Gemini Observatory, NSF's NOIRLab, 670 N A'ohoku Place, Hilo, Hawai'i 96720, USA
             }

   \date{Received XXX; accepted YYY}

% \abstract{}{}{}{}{} 
% 5 {} token are mandatory
 
  \abstract
  % context heading (optional)
  % {} leave it empty if necessary  
   {We present VLT/MUSE observations targeting the extended Lyman-$\alpha$ (\lya) emission of five high-redshift (z$\sim$3-4) submillimeter galaxies (SMGs) with increasing quasar (QSO) radiation: two SMGs, two SMGs hosting a QSO, and one SMG hosting a QSO with a SMG companion (QSO+SMG). 
    These sources should be 
    located in dark matter halos of comparable masses (average mass of $M_{\rm DM}\sim10^{12.2}$\,\msun). We quantify the luminosity and extent of the \lya emission, together with its kinematics, and examine four \lya powering mechanisms: photoionization from QSOs/star formation, shocks by galactic/QSO outflows, gravitational cooling radiation, and \lya photons resonant scattering. We find a variety of \lya luminosities and extents, with the QSO+SMG system displaying the most extended and bright nebula, followed by the SMGs hosting a QSO, and finally the undetected circumgalactic medium (CGM) of SMGs.
    This diversity implies that gravitational cooling is unlikely to be the main powering mechanism. We show that photoionization from the QSO and QSO outflows can contribute to power the emission for average densities $n_{\rm H}>0.5$~cm$^{-3}$. Moreover, the observed \lya luminosities scale with the QSO’s budget of \lya photons modulo the dust content in each galaxy, highlighting a possible contribution from resonant scattering of QSO's radiation in powering the nebulae. We find larger \lya linewidths (FWHM$\gtrsim1200$\,km\,s$^{-1}$) than usually reported around radio-quiet systems, pointing to large-scale outflows. A statistical survey targeting similar high-redshift massive systems with known host properties is needed to confirm our findings.} 
    

   \maketitle

\section{Introduction}\label{sec:intro}

In the $\Lambda$CDM paradigm, galaxies form and evolve embedded in a filamentary cosmic web, and grow by accretion of material from their surrounding medium. In turn, galaxies are expected to modify and pollute their vicinities through several processes (e.g., photoionization, outflows). The gas interface, extending beyond a galaxy's interstellar medium, but bound to the galaxy's halo, is often called the circumgalactic medium (CGM; \citealp[e.g.,][]{Tumlinson2017}). The CGM extends over hundreds of kpc and naturally encodes information on the interactions between a galaxy and its surroundings (e.g., inflows, outflows), making its direct study fundamental for the understanding of galaxy formation and evolution.

The current generation of sensitive integral-field unit (IFU) spectrographs, such as the Multi Unit Spectroscopic Explorer \citep[MUSE;][]{Bacon2010} at the Very Large Telescope (VLT) and the Keck Cosmic Web Imager \citep[KCWI;][]{Morrisey2018} at the Keck Observatory, are able to push observations to unprecedented surface brightness limits (${\rm SB~\sim~10^{-19}\,erg\,s^{-1}\,cm^{-2}\,arcsec^{-2}}$) allowing us to study the CGM in emission. For example, observations of the Hydrogen Lyman-$\alpha$ (\lya) 
emission line surrounding high-z (${\rm 2<z<6}$) quasars are now routinely reported \citep[e.g.,][]{Borisova2016, FAB2018, FAB2019, Farina2019, Cai2019, O'Sullivan2020, Mackenzie2020, Fossati2021}. The \lya emission traces large-scale (${\rm \sim100\,kpc}$), cool (${\rm T\sim10^4\,K}$), and massive 
gas reservoirs within the dark-matter halos hosting quasars at these redshfits, which are expected to have masses in the range ${\rm M\sim10^{12}-10^{13}\,\msun}$ \citep[e.g.,][and references therein]{Timlin2018}. 

The main powering mechanism of the extended \lya glow is frequently invoked 
as photoionization from the associated bright quasar followed by recombination in optically thin gas 
\citep[e.g.,][]{Heckman1991, Cantalupo2014, Hennawi2015,FAB2015b,Cai2018,FAB2018}. However, the detailed radiative transfer of this resonant line emission, and the balance between different powering mechanisms is still unclear and debated. Hydrodynamical simulations of $z\sim2$ halos show that photoionization from the central active galactic nucleus (AGN) and star formation from companion galaxies can produce extended \lya emission \citep[][]{Gronke2017}. Also, 
recent simulations of $z\gtrsim 6$ quasar nebulae showed that scattering of Ly$\alpha$ photons is required to explain the morphology of the extended Ly$\alpha$ emission \citep[][]{Costa2022}. This work also showed that Ly$\alpha$ photons from the quasar's broad-line-regions (BLR), added in post-processing, can contribute to the powering of the extended emission as quasar feedback is able to open channels of least resistance for the propagation of such photons out to CGM scales. 

There are at least four possible mechanisms that could act together to power the extended \lya emission: 
(i) photoionization from the central AGN or companion galaxies, (ii) shocks powered by galactic/AGN outflows, (iii) gravitational cooling radiation, and (iv) resonant scattering of \lya. Disentangling their relative roles and making an unique interpretation of observations is challenging \citep[e.g.,][]{FAB2015a, Mackenzie2020}, and it is at the root of the difficulties in firmly constraining the physical properties of the emitting gas. Possible avenues to assess the contribution of different powering mechanisms are polarimetric observations of the Ly$\alpha$ emission \citep[e.g.,][]{Hayes2011,Kim2020}, or constraints on additional emission lines such as H$\alpha$, \ion{C}{iv}$\lambda1549$\footnote{The \ion{C}{iv} line is a doublet with wavelengths $1548$\,\AA\ and $1550$\,\AA.} and \ion{He}{ii}$\lambda1640$, which also aid in determining the ionization state and metallicity of the CGM \citep[e.g.,][]{FAB2015b}.
For each of the mechanism, the following expectations hold:

\begin{itemize}[leftmargin=.2in]
    \item Photoionization from the central AGN: The AGN illuminates the CGM, which in turn reprocesses the UV emission into a recombination cascade including detectable \lya, H$\alpha$ and \ion{He}{ii} emission \citep[e.g.,][]{Heckman1991b, Christensen2006, Smith2009, Geach2009, Humphrey2013}. If the CGM is already enriched (e.g., from outflows) then extended \ion{C}{iv} emission could also be detected.
    \item Shocks powered by galactic/AGN outflows: The \lya linewidths observed for nebulae around quasars are generally as expected for quiescent gas in virial equilibrium with the surrounding dark matter halo \cite[Full Width Half Maximum ${\rm FWHM}\sim600{\rm \,km\,s}^{-1}$; e.g.,][]{FAB2019,Cai2019,O'Sullivan2020,Lau2022}. However, broader linewidths on tens of kpc around some objects \citep[e.g.,][]{Ginolfi2018, Vidal-Garcia2021} are evidence for larger turbulence likely caused by outflows from the central AGN. Shocks produced by this outflowing material can produce \lya emission through collisional excitation and ionization \citep[e.g.,][]{Taniguchi&Shioya2000, Taniguchi2001, Ohyama2003, Wilman2005, Mori&Umemura2006}. The same mechanisms could also allow for the emission of \ion{C}{iv} and \ion{He}{ii} lines.
    \item Gravitational cooling radiation: 
    As gas cools within dark matter halos, it will radiate away the lost gravitational potential energy through cooling channels. 
    The main result is the emission of \lya photons produced by collisional excitations \citep[e.g., ][]{Haiman2000, Furlanetto2005, Dijkstra2006, Faucher-Giguere2010, Rosdahl&Blaizot2012}. Detecting any other line emission due to this mechanism would be difficult with current facilities. 
    \item Resonant scattering of \lya photons originating from compact sources: 
    \lya photons produced by the QSO, the host galaxy and companion galaxies can contribute to an extended \lya glow as they undergo resonant scattering while propagating outwards through the CGM \citep[e.g.,][]{Dijkstra&Loeb2008, Heyes2011, Cen&Zheng2013, Cantalupo2014}. \ion{C}{iv} is also a resonant transition and it could be detected as extended glow if scattered by a C$^{3+}$-rich medium. 
    Recombination transitions (e.g.,H$\alpha$, \ion{He}{ii}) are expected to be more compact and narrower than \lya  
    \citep[e.g.,][]{Prescott2015}.
\end{itemize}

In addition, resonant scattering is expected to take place also for the \lya photons originating from the first three mechanisms. This results in broader lines and larger nebulae due to the diffusion in space and frequency of the \lya photons (e.g., \citealt{Dijkstra2019}).

Stacking analysis of several objects start to reveal extended line emission besides Ly$\alpha$. Using MUSE data on 80 $z\sim3$ QSOs (1 hour/source) \citet{Guo2020} reported average \ion{C}{iv}, \ion{He}{ii}, and \ion{C}{iii$\left.\right]$} SB profiles, implying high metallicities in the inner CGM. Moreover, 
the latest and deepest (4 hours/source with MUSE) survey targeting high-redshift quasars was able to reveal extended emission from \lya, \ion{C}{iv} and tentatively from \ion{He}{ii} (barely $2\sigma$ significance) in their stacked analysis of 27 targets \citealp[][]{Fossati2021}. 
They derived line ratios and highlighted the difficulties in modeling the observed metallicities, but conclude that the CGM gas is metal polluted at redshift $z=3-4.5$. In particular, they discussed that resonant scattering of quasar's \lya photons could be an explanation to the observed ratios and to why their sample shows emission which is more extended in resonant lines (\lya and \ion{C}{iv}) than non-resonant lines (\ion{He}{ii}).

The longer term goal of our endeavor 
is 
to break the degeneracies between different powering mechanisms. In this  
project, we present the \lya extended emission around different types of galaxies: 
unobscured quasars with a companion submillimeter galaxy \cite[SMG; e.g., ][]{Smail1997}, SMGs hosting quasars, and SMGs. While all these objects are predicted to reside in similarly massive dark matter halos as quasars (${\rm 10^{12-13}\,\msun}$; \citealp[e.g.,][]{Wilkinson2017, Garcia-Vergara2020,Lim2020}), and to have large star-formation rates, they have different contributions of quasar radiation (from none to bright unobscured quasars). We stress that the clustering of quasars is independent of the luminosity of the quasar ($-28.7 < M_{\rm i} < -23.8$, \citealt[][]{Eftekharzadeh2015}), hence looking at different luminosity systems probes the effect of different levels of ionizing radiation in the same environment. Therefore, once corrected for the cosmological surface brightness dimming, we expect that these systems (i) have similar contributions to the total \lya emission radiated from gravitational cooling 
\citep[][]{Haiman2000}, (ii) have different contributions from AGN photoionization, star-formation, and from \lya resonant scattering, proportional to the AGN luminosity, star-formation activity, and \lya photon budget in compact sources, respectively, and (iii) have a shock contribution only when evidence of violent kinematics are present on CGM scales.  

We emphasize that all our targets have large infrared luminosities (${\rm L_{FIR}\sim10^{12-13}\,\lsun}$), similar to those of SMGs, and which are caused by dust heated by high rates of obscured star formation, with star formation rates of ${\rm SFR}\sim100-1000\,\msun$\,yr$^{-1}$ \citep{Blain1999, Casey2014}. In particular, recent works have shown that the distribution of dust in some SMGs appears clumpy and displaced from UV regions, which allows a fraction of their UV photons to escape \citep[e.g.,][]{Hodge2015, Chen2017}. Our observations could therefore estimate what fraction of escaping UV photons from star formation impinge on their CGM. In other words, if cool gas is present around these sources, then we expect that extended \lya emission due to star formation might be detected when there are enough escaping UV photons reaching the surrounding gas distribution.

This paper is organized as follows. In Section~\ref{sec:data}, we describe in detail the sample, observations and data reduction. In Section~\ref{sec:analysis}, we describe the analysis carried out in order to reveal the extended emission around our sources. In Section~\ref{sec:results}, we present our results, discuss the properties of the detected extended \lya emission such as brightness, morphology and kinematics, and report constraints on additional emission lines. 
In Section~\ref{sec:discussion} we compare  
our observational results with previous literature and discuss 
the \lya powering mechanisms mentioned above, showing that the firmest constraint is that against the gravitational cooling scenario.  above. 
We summarize our findings in Section~\ref{sec:summary}.
We adopt a flat $\Lambda$CDM cosmology with $H_0=70\,\kms\,\Mpc^{-1}$, $\Omega_{\rm m}=0.3$ and $\Omega_{\rm \Lambda}=0.7$, for which the scale of one arcsecond corresponds to $\rm{7.0\,kpc}$
at the mean redshift of our sources ($\rm{z\sim4}$).

%--------------------------------------------------------------------
\section{Observations and Data Reduction} \label{sec:data}

\subsection{Sample selection}
This  
project exploits the spectral-imaging capabilities of VLT/MUSE to target five systems with strong star formation, including a SMG hosting a QSO with an SMG companion (QSO+SMG), two SMGs hosting a QSO, and two SMGs.  
In particular, the SMGs are part of the first samples with a firm estimate of their systemic redshifts through the observation of molecular tracers. Our sample is composed as follows:

\begin{itemize}[leftmargin=.2in]
    \item One radio-quiet unobscured bright quasar with a submillimeter companion galaxy $\sim$4\arcsec\ to the NW of the QSO, ${\rm BR1202\textsc{-}0725}$ \citep[e.g.,][]{Omont1996b, Drake2020};
    \item Two quasars in SMGs, ${\rm G09\textsc{-}0902\textsc{+}0101}$ and ${\rm G15\textsc{-}1444\textsc{-}0044}$ \citep[][]{Fu2017}. Even though these sources are bright in the submillimeter, their quasar spectra appear unobscured in the rest-frame UV along the line of sight. They are therefore unobscured QSOs hosted by SMGs. They are about one order of magnitude fainter than the quasar in BR1202-0725;
    \item Two isolated SMGs (with no AGN) with spectroscopic redshift from [CII] and CO, ${\rm ALESS61.1}$ and ${\rm ALESS65.1}$ \citep[][]{Swinbank2012,Birkin2021}.
\end{itemize}

\begin{figure*}
    \centering
    \includegraphics[height=0.7\textheight]{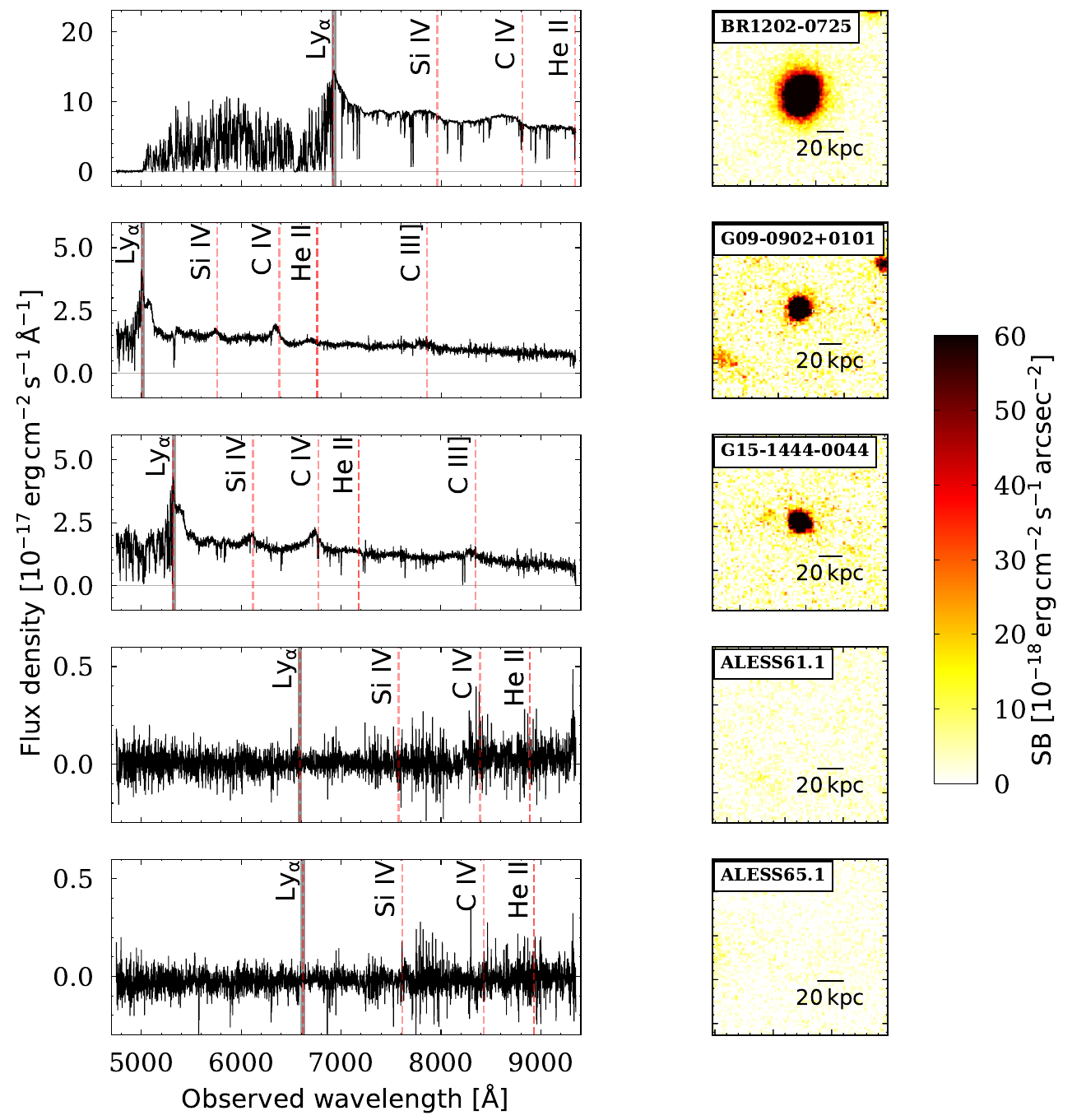}
    \caption{Overview of the five targeted systems before subtraction of the unresolved quasar's point spread function and continuum sources (see Section~\ref{sec:analysis}). {\it Left:} Integrated spectra of the QSOs and SMGs inside a 3\arcsec radius aperture. Vertical dashed lines indicate the positions of the \lya and quasar broad lines (\ion{Si}{iv}, \ion{C}{iv}, \ion{He}{ii} and \ion{C}{iii$\left.\right]$}). The gray shaded region indicates the FWHM of the nebular \lya emission for the quasars (see Figure~\ref{fig:sbmaps-qso} and Table~\ref{table:SFRv2}) and the 30\AA\ narrow band used for the SMGs (see Figure~\ref{fig:sbmaps-smg}). 
    {\it Right:} 30\,\AA\ Surface brightness maps at the expected \lya wavelength of the targeted QSOs and SMGs before PSF- and continuum-subtraction (see Section~\ref{sec:analysis}). The maps have a side of 20\arcsec, corresponding to about 129, 152, 148, 133 and 132\,kpc for $\rm{BR1202\textsc{-}0725}$, $\rm{G09\textsc{-}0902\textsc{+}0101}$, $\rm{G15\textsc{-}1444\textsc{-}0044}$, ${\rm ALESS61.1}$ and ${\rm ALESS65.1}$ respectively.}
    \label{fig:original_spectra}
\end{figure*}

\begin{table*}
\centering
\caption{Physical properties of the targeted sample and observing log.}
\resizebox{\textwidth}{!}{
\begin{tabular}{|c|c|c|c|c|c|c|c|c|c|c|}
 \hline
 ID & Object & Redshift$^{\rm a}$ & $\log{L_{\rm bol}^{\rm QSO}/\lsun}^{\rm b}$ & $\log{L_{\rm IR}^{\rm SF}/\lsun}^{\rm c}$ & ${\rm SFR}^{\rm d}$ & $M_{\rm dust}^{\rm e}$ & Seeing$^{\rm f}$ & Exp. time & ${\rm SBlim_{Ly\alpha}}$ & 
 Type\\
 & & & & & $[{\rm \msun\,yr^{-1}}]$ & [$10^8$\,\msun] & ${\rm [arcsec]}$ & [minutes] & $[{\rm \times10^{-18}]^g}$ 
 & \\
 \hline
 1 & ${\rm BR1202\textsc{-}0725}$ & $4.6942\pm0.0003$ ([\ion{C}{ii}]) & 47.7 & $13.7\pm0.2$
 & $5000\pm2300$
 & $23.6\pm3.1$ & 1.1 & 126 &
 $3.2^{\rm i}$ 
 & QSO+SMG\\
 2 & $\rm{G09\textsc{-}0902\textsc{+}0101}$ & 3.120 & 46.5 & $13.2\pm0.1$ & 
 $1600\pm370$
 & $17.4\pm3.6$ & 1.0
 & 44 &
 $3.1$ 
 & SMG hosting a QSO$^{\rm h}$\\
 3 & $\rm{G15\textsc{-}1444\textsc{-}0044}$ & 3.375 & 46.6 & $13.3\pm0.1$ & 
 $2000\pm460$
 & $19.7\pm2.9$ & 1.0 & 44 & 3.3
 & SMG hosting a QSO$^{\rm h}$\\
 4 & ALESS61.1 & $4.4189\pm0.0004$ (CO(4-3))& -- & $12.6\pm0.1$ 
 & $400\pm90$
 & $4.8\pm0.9$ & 1.3 & 86 &
 $2.5$ 
 & SMG\\
 5 & ALESS65.1 & $4.4445\pm0.0005$ (CO(4-3))& -- & $12.6\pm0.1$
 & $400\pm90$ 
 & $4.7\pm1.1$ & $1.2$
 & 86 &
 $2.2$ 
 & SMG\\
 \hline
\end{tabular}
} 
\flushleft{$^{\rm a}$ Redshifts from \citet{Carilli2013}, \citet{Fu2017} and \citet{Birkin2021}.
In brackets we report the line used for precise redshift estimates. For G09-0902+0101 and G15-1444-0044, the redshifts are the best values from \citet{Lyke2020}, whose uncertainties are of the order of few hundreds of km\,s$^{-1}$.\\
$^{\rm b}$ Bolometric luminosities computed from the observed luminosities at 1450\AA\ using the relation by \citet{Runnoe2012}.\\
$^{\rm c}$ IR luminosities at $8-1000$\,$\mu$m due to star formation re-computed in this work, and consistent within uncertainties with values in \citet{Wagg2012}, \citet{Fu2017} and \citet{Birkin2021}. For BR1202-0725 we only use the datapoints for the quasar from \citet{Omont1996a} and \citet{Iono2006}.\\
$^{\rm d}$ SFR computed using the formula ${\rm SFR}=[L_{\rm IR}^{\rm SF}/(10^{10}\,{\rm L}_{\odot})]\,\msun\,{\rm yr}^{-1}$ (\citealt{Fu2017}). The errors reflect only the uncertainty on the IR luminosities.\\
$^{\rm e}$ Dust masses computed assuming a modified blackbody with dust emissivity spectral index $\beta=2.0$ and fixing the dust temperature only for BR1202-0725 as we only use two datapoints not covering the SED peak ($T_{\rm dust}=68$\,K; \citealt{Leech2001}). For the other sources we found $T_{\rm dust}\sim55$\,K.\\
$^{\rm f}$ The seeing is estimated by fitting a 2D Moffat model to point sources in each field (see Appendix~\ref{sec:appendix-seeing}).\\

$^{\rm g}$ $2\sigma$ surface brightness detection level within 1 square arcsecond and $30$~\AA\, at the expected Ly$\alpha$ wavelengths, in units of ${\rm erg\,s^{-1}\,cm^{-2}\,arcsec^{-2}}$.\\

$^{\rm h}$ The host-galaxies are submillimeter bright with the flux at 870\,$\mu$m of $S_{870} = 14.4$ and $6.0$\,mJy for G09-0902+0101 and G15-1444-0044, respectively (\citealt{Fu2017}).\\

$^{\rm i}$ At the expected Ly$\alpha$ wavelength for BR1202-0725 there are brighter sky lines than for all other sources. This is why the ${\rm SB}_{\rm lim}^{\rm Ly\alpha}$ for BR1202-0725 is comparable to the other sources despite the longer exposure time.}
\label{tab:Data}
\end{table*}

\subsection{VLT/MUSE observations and data reduction}

The VLT/MUSE data were taken as part of ESO programmes 0103.A-0296(A) and 0102.A-0403(A) in service modes on UT dates between January 2019 and August 2019 (PI: F. Arrigoni Battaia). The observations were taken in wide field mode, resulting in a $0.2\arcsec$ spatial sampling of a $1^{\prime}\times1^{\prime}$ field of view. In this configuration MUSE covers the spectral range 4750 -- 9350\,\AA\ with a spectral resolution of $R \sim 2230$ at 6078~\AA\ (the Ly$\alpha$ wavelength at the mean $z$ of the sample). The observing blocks were organized in three exposures of about 15 minutes each with $<5\arcsec$ shifts and $90$ degree rotations. The observing conditions were clear and the seeing at the expected \lya wavelength was of $1.13\arcsec$, on average. In addition, for ${\rm BR1202\textsc{-}0725}$ we include, re-reduce, and add to our dataset the 48 minutes dataset from \citet{Drake2020}, which was taken with MUSE in the same configuration as part of ESO programme ${\rm 0102.A-0428(A)}$ (PI: E.~P. Farina), with two exposures with a $<5\arcsec$ shift and a $90$ degree rotation. These observations were obtained under good weather conditions and a 0.6\arcsec seeing. 
In Table~\ref{tab:Data} we list the total exposure time and the seeing at the \lya wavelength measured for each field from its final datacube (see Appendix~\ref{sec:appendix-seeing}).

The data reduction was performed using the MUSE pipeline version 2.8.3 \citet{Weilbacher2012,Weilbacher2014,Weilbacher2020} as described in \citet{Farina2019}, which consist of bias and dark subtraction, flat correction, wavelength calibration, illumination correction and standard star calibration. We continue with the removal of sky emission residuals, which we perform using the Zurich Atmospheric Purge software (ZAP; \citealt{Soto2016}). The reduced datacubes contain both the science data cubes and the associated variance cubes. The later are rescaled as usually done in the literature to reflect the empirical variance in the data cubes, since the pipeline underestimates the variance due to correlated noise at pixel level (e.g., \citealt{Bacon2015}). The scaling factor is found to be on average 1.4, consistent with previous works \citep[e.g.,][]{Borisova2016}, and is applied layer by layer to the variance cube. The final variance cubes are used to obtain surface brightness (SB) limits and associated errors of our estimates.
After data reduction, we manually mask artifacts that appear due to the edges of the MUSE IFUs. This process needs to be performed for each single exposure before combining. During this step, about $\sim4\%$ of the pixels are masked on average per exposure. Masking is done by inspection of white images (computed by collapsing the datacube along the spectral axis using the full wavelength range). 
After masking, each exposure is aligned by performing a 2D gaussian centroid fitting of point sources (stars) in each field for reference, then a simple offset is computed. Finally the datacubes are combined using the median, the variance cubes are also combined by performing error propagation.

At the expected wavelength of the \lya emission, the final $2\sigma$ SB limit is on average ${\rm \sim3\times10^{-18}\,erg\,s^{-1}\,cm^{-2}\,arcsec^{-2}}$ for an aperture of 1 square arcsecond and a ${\rm 30\,\angstrom}$ narrow band. We list in Table~\ref{tab:Data} these SB limits.

Finally, Figure~\ref{fig:original_spectra} shows the integrated spectra obtained from our MUSE data inside a 3\arcsec radius aperture for each of the systems studied here, together with SB maps of 30\,\AA\ narrow-bands centered at the expected \lya line wavelength before subtracting the quasar's point spread function and continuum sources. We can already see from this figure that there is no detected emission from the SMGs. Note that for BR1202-0725, the observed \ion{He}{ii} wavelength is located right before the edge of the spectral range (about ${\rm 9\,\angstrom}$). In Section~\ref{sec:analysis} we discuss in detail these observations and their analysis.

\subsection{Physical properties of the sample}
\label{subsec:physProp}

To homogenize our sample, we re-compute some of the galaxies' physical properties in a consistent way for all sources, namely infrared (IR) luminosity, SFRs and dust masses. To do this, we use ancillary submillimeter data.
We construct the 8-1000\,$\mu$m spectral energy distribution using the datapoints from \citet[][]{Omont1996a} (at $1250$~$\mu$m with IRAM 30\,m) and \citet[][]{Iono2006} (at $900$~$\mu$m with the Submillimeter Array) for BR1202-0725, from \citet[][]{Fu2017} (at 250, 350 and 500\,$\mu$m with Herschel/SPIRE and 870\,$\mu$m with ALMA) for G09-0902+0101 and G15-1444-0044, and from \citet[][]{Swinbank2012} (at 250, 350 and 500\,$\mu$m with Herschel/SPIRE and 870\,$\mu$m with ALMA) for ALESS61.1 and ALESS65.1. In Table~\ref{tab:ancillary} we summarize the fluxes of these submillimeter observations.

\begin{table}
\centering
\caption{Summary of the ancillary data fluxes used to estimate the physical properties of the sample.}
\resizebox{1.0\columnwidth}{!}{
\begin{tabular}{|c|c|c|c|c|c|c|}
\hline
\hline
ID & $S_{250}$ & $S_{350}$ & $S_{500}$ & $S_{870}$ & $S_{900}$ & $S_{1250}$ \\
   & mJy       & mJy       & mJy       & mJy       & mJy       & mJy       \\ 
\hline 
\hline

1  &--      &   --   &  --    & --     &   $32\pm4$   &     $12.59\pm2.28$       \\
2  &   $53.7\pm7.5$   &   $56.7\pm8.6$   &   $45.9\pm9.3$   &   $14.4\pm0.8$   &   --   &     --       \\
3  &   $47.3\pm6.6$   &  $61.2\pm8.2$    &   $58.8\pm8.8$   &   $6.0\pm0.6$   &  --    &    --        \\
4  &   $4.3\pm1.5$   &  $7.4\pm1.6$    &  $10.2\pm1.7$    &   $4.32\pm0.44$   &   --   &    --        \\
5  &   --  &    $7.6\pm1.4$    &  $10.2\pm1.5$    &  $4.24\pm0.49$    &   --   &     --       \\
\hline
\end{tabular}
} 

\label{tab:ancillary}
\end{table}

 We estimate the dust masses of each source assuming a modified black body of the form ${S_{\nu_0} \propto B_{\nu}\times[1-\exp{(- \nu / \nu_0 )^\beta}]}$ with $\nu_0=2.0$\,THz and dust emissivity spectral index $\beta=2.0$ \citep[][]{Fu2017}, integrating to obtain the total IR luminosity between $\lambda=8-1000\,\mu$m. We then use this integrated IR luminosity to estimate the SFR for each source using the formula ${\rm SFR}=[L_{\rm IR}^{\rm SF}/(10^{10}\,{\rm L}_{\odot})]\,\msun\,{\rm yr}^{-1}$ from \citet[][]{Fu2017}, which assumes a \citet[][]{Chabrier2003} initial mass function. This procedure is reasonable as it has been shown that most of the IR emission in these sources is dominated by star formation in the host galaxy (e.g., \citealt{Iono2006,Fu2017}).
The dust masses are derived using equation (3) of \citet[][]{Chen2021}, assuming a rest-frame dust mass absorption coefficient $\kappa_{\rm 850\,\mu m}=0.431\,{\rm cm^2\,g^{-1}}$ from \citet[][]{Li2001}. For BR1202-0725 we fix the dust temperature to $T_{\rm dust}=68$\,K \citep[][]{Leech2001} as we only use the two available datapoints for the quasar, which are not covering the spectral energy distribution (SED) peak. For the other sources we find $T_{\rm dust}\sim55$\,K. If we adopt 
this same temperature for BR1202-0725 we derive a 20\% higher dust mass.
The derived dust masses (see Table~\ref{tab:Data}) are consistent with dust masses reported in the literature for similar objects. 
Specifically, the dust masses of ALESS61.1 and ALESS65.1 are in agreement with the values estimated by \citet{Birkin2021}.

Further, we provide a rough calculation to show  
that the  
halo masses of these systems are within the range expected for quasars and SMGs at these redshifts ($10^{12}-10^{13}\,\msun$). Specifically, we predict their DM halo masses using the halo mass-stellar mass relation by \citet{Moster2018} at their redshifts. We stress that the obtained estimates are affected by large uncertainties given the scatter in this relation. For the two SMGs, ALESS61.1 and ALESS65.1, we can rely on their stellar masses obtained through SED fitting by \citet{Birkin2021},  ${\rm log(}M_{*}/[\rm M_{\odot}])=10.33^{+0.18}_{-0.01}$ and ${\rm log(}M_{*}/[\rm M_{\odot}])=10.48^{+0.19}_{-0.13}$, respectively. 
For the three quasar systems, we instead convert their total gas mass to a stellar mass, assuming the gas to stellar mass ratio at their redshift from \citet{Birkin2021} (e.g., see their Figure 9, left panel). For this step, we homogenize all the measurements obtaining gas masses from the far-IR continuum following equation 3 in \citet{Tacconi2020}, assuming a gas-to-dust ratio of 100 (e.g., \citealt{Riechers2013}). We find $M_{\rm gas }=8\times10^{10}\,\msun$ for BR1202-0725, and 
$M_{\rm gas }=6.3\times10^{10}\,\msun$ and $M_{\rm gas }=1.9\times10^{10}\,\msun$ for G09-0902+0101 and G15-1444-0044, respectively. The value for BR1202-0725 is consistent with the value computed from CO emission and assuming an $\alpha_{\rm CO}=0.8$\,M$_{\odot}$\,K$^{-1}$\,km\,s$^{-1}$\,pc$^2$ \citep[][]{Riechers2006}.
We therefore homogenized the values in \citet[][]{Birkin2021} to this $\alpha_{\rm CO}$. We stress that they targeted 61 SMGs with no clear sign of AGN activity, therefore the used ratio may introduce a systematic bias for our QSOs which are hosted by SMGs. 
We find stellar masses of $M_{*}=6.7^{+5.9}_{-2.1}\times10^{10}\,\msun$ for BR1202-0725, and 
$M_{*}=8.0^{+10.9}_{-2.8}\times10^{10}\,\msun$ and $M_{*}=2.3^{+2.7}_{-0.8}\times10^{10}\,\msun$ for G09-0902+0101 and G15-1444-0044, respectively. 
Therefore, the expected DM halo masses using the halo mass-stellar mass relation by \citet{Moster2018} for the targeted sources are: ${\rm log(}M_{\rm DM}/[\rm M_{\odot}])\boldsymbol{\sim}{12.3}$  
for BR1202-0725, 
${\rm log(}M_{\rm DM}/[\rm M_{\odot}])\boldsymbol{\sim}{12.5}$ 
and ${\rm log(}M_{\rm DM}/[\rm M_{\odot}])\boldsymbol{\sim}{11.9}$ 
for G09-0902+0101 and G15-1444-0044, 
and ${\rm log(}M_{\rm DM}/[\rm M_{\odot}])\boldsymbol{\sim}{11.9}$ 
and ${\rm log(}M_{\rm DM}/[\rm M_{\odot}])\boldsymbol{\sim}{12.0}$ 
for ALESS61.1, and ALESS65.1, respectively. 
We quote the halo mass estimates as approximate because of the large uncertainties inherent to the use of the relation in \citet{Moster2018}, the use of gas masses to obtain stellar masses, and the  
uncertainty on the $\alpha_{\rm CO}$ assumed, which could be a factor of 5 higher if $\alpha_{\rm CO}$ is close to the galactic value \citep[][]{Bolatto2013}. Notwithstanding these uncertainties, the values we find 
are consistent with the range of halo masses from clustering studies (\citealp[][]{Timlin2018}, and references therein). These estimates are key in placing these systems in the context of galaxy formation, and in quantifying the expected signal for gravitational cooling.
Table~\ref{tab:Data} summarizes most of the properties of our targets (redshift, quasar's bolometric luminosity, infrared luminosity due to star formation, SFR, estimated dust mass, seeing at the \lya wavelenth, exposure time and their type) with their respective references.

%%%%%%%%%%%%%%%%%%%%%
%      ANALYSIS     %
%%%%%%%%%%%%%%%%%%%%%

\section{Analysis: Revealing the large-scale emission}\label{sec:analysis}

While SMGs are faint in rest-frame UV, we need to remove the unresolved quasar's point spread function (PSF) to reveal the extended emission around them. In the following subsections we describe the method used to reveal the extended \lya emission. For this, we develop new Python custom routines that construct and subtract the empirical quasar's PSF, similarly to what is usually done in the literature \citep[e.g.,][]{Borisova2016, Husemann2018, Farina2019, O'Sullivan2020}. In summary, we build a wavelength-dependant empirical quasar's PSF which is later subtracted from the datacubes, we then subtract continuum sources in order to obtain datacubes with pure extended line emission. The steps for this method are described below.

\subsection{PSF Subtraction}\label{subsec:psfsub}

The quasars in our sample all outshine the host galaxy at the wavelengths targeted by MUSE. As usually done in the literature, the PSF is computed at each wavelength by constructing pseudo narrow-band images of 150 channels ($187\,\angstrom$) centered at the quasar's position, and inside a box of size 6 times the seeing to ensure that its contribution at the largest radii is negligible. We describe how the seeing is estimated for each observation in Appendix~\ref{sec:appendix-seeing}.  

We require that the PSF is not computed where we expect the extended line emission, corresponding to wavelength ranges affected by broad quasar lines (such as, e.g., \lya, CIV, HeII). These wavelength ranges are neglected and their PSF is computed at the first available redshifted 
pseudo narrow-band.
To identify where these wavelength ranges are, we design a custom routine that automatically detects the location of line emission as explained in detail in Appendix~\ref{sec:lines_identification}. In summary, we find the wavelength of local peaks in the spectra that indicate expected line emission and define a conservative spectral window of about $\Delta \lambda=430-560\,$\AA\ ($\Delta v=24400-26000\,{\rm km\,s^{-1}}$) centered on each line, which is excluded from computing the PSF. Such velocity ranges are much larger than the width of the discovered emission (Section~\ref{sec:results}). 

The empirical wavelength-dependent PSF is then rescaled at each layer to match the quasar's flux within the central 1\,${\rm arcsec^2}$. Due to possible cosmic rays, the scaling factor is computed from the average sigma clip of the normalization region. We subtract the rescaled PSF from the original datacube layer by layer, after masking negative pixels (at the edges of the PSF where it approaches the background) in order to not introduce spurious signal. In each field, we finally mask the 1\,${\rm arcsec^2}$ normalization region and exclude it from our analysis.

The resulting PSF subtracted datacubes are already able to highlight the extended emission around the quasars. However continuum sources in the field still need to be removed in order to not contaminate the extended emission.

\subsection{Continuum subtraction}\label{ssubsec:contsub}

After subtracting the empirical PSF, we remove the contribution from continuum sources in all spaxels of our datacubes. For this, we first create a sky mask using the PSF-subtracted white light image, for which we estimate the noise floor 
using an average sigma clip algorithm and identify sources above 5 times this threshold. We estimate the continuum of the identified sources by smoothing their spectra at each spaxel using an order 5 polynomium, as done in the MPDAF Python package from \citet[][]{Bacon2016}. Finally we subtract the smoothed spectrum from each voxel of the identified sources. This process results in a datacube of only line emission. To check the robustness of our PSF and continuum subtractions we compare with the results from \citet[][]{Drake2020} for BR1202-0725 and find consistent results for the \lya emission (emission morphology and levels), though our data are deeper.

%%%%%%%%%%%%%%%%%%
%     RESULTS    %
%%%%%%%%%%%%%%%%%%

\section{Results}\label{sec:results}

The PSF- and continuum-subtracted cubes have been used to extract line emission nebulae associated with the five targets. Below, we focus on the surface brightness maps, integrated spectra, radial profiles and kinematics of the \lya line emission. We also provide constraints on 
additional rest-frame UV line emission,  
\ion{C}{iv}, \ion{Si}{iv}, \ion{He}{ii}, and \ion{C}{iii}]\footnote{The \ion{C}{iii}] transition is a doublet of a forbidden and semi-forbidden transitions with wavelengths 1907 and 1909~\AA, respectively.}, when available within the MUSE wavelength range. 

\subsection[Ly\texorpdfstring{$\alpha$}{alpha} surface brightness maps and nebula spectra]{Ly$ \alpha$ surface brightness maps and nebula spectra}\label{subsec:sbmaps}

\begin{figure}
    \centering
    \includegraphics[width=0.45\textwidth]{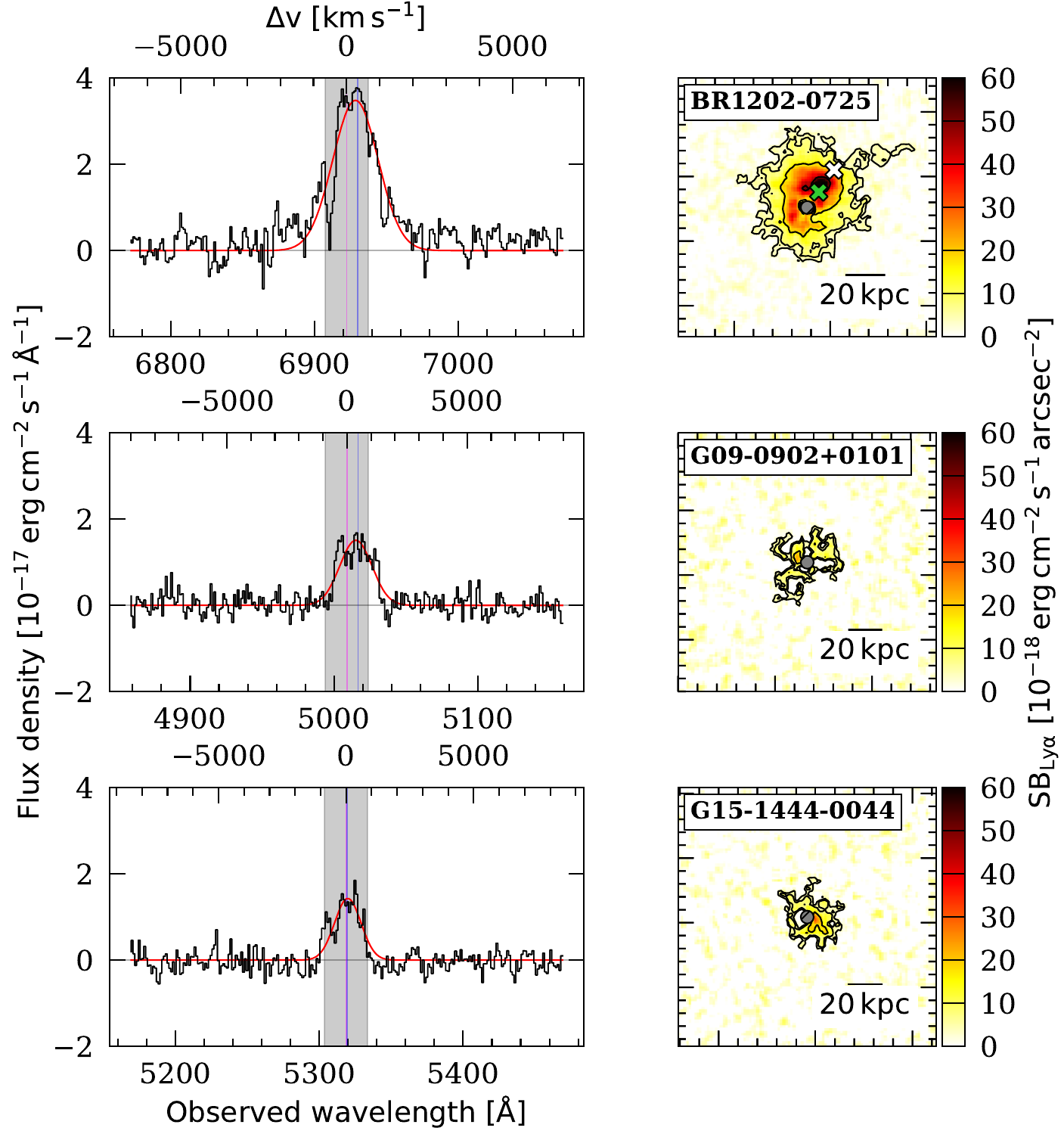}
    \caption{Integrated spectra and surface brightness maps of the extended \lya emission around the three systems with AGN. \textit{Left:} Spectra of the identified \lya nebulae (black curve), integrated inside the $2\sigma$ contour. The gray shaded area shows the channels used to build the surface brightness map. The magenta vertical line indicates the \lya wavelength from the systemic redshift of the quasars, while the blue vertical line indicates the wavelength at the peak of the \lya spectra computed from the first moment of the line. Overlayed in red is a 
    Gaussian fit to the spectra. The velocity shift and the FWHM of each system are listed in Table~\ref{table:SFRv2}. 
    The top axis of the spectra panels show the velocity shift with respect to the systemic redshift in $\kms$. \textit{Right:} \lya surface brightness maps of $\rm{BR1202\textsc{-}0725}$ (top), $\rm{G09\textsc{-}0902\textsc{+}0101}$ (middle) and $\rm{G15\textsc{-}1444\textsc{-}0044}$ (bottom) at the expected wavelength from their systemic redshift (Table~\ref{tab:Data}). The maps have a side of 20\arcsec, corresponding to about 129, 152 and 148\,kpc for $\rm{BR1202\textsc{-}0725}$, $\rm{G09\textsc{-}0902\textsc{+}0101}$ and $\rm{G15\textsc{-}1444\textsc{-}0044}$ respectively. A scalebar of 20\,kpc is indicated on each map. The black contours indicate surface brightness levels of $[2, 4, 10, 20, 50]$ $\sigma$ (see Table~\ref{tab:Data}). 
    These maps are smoothed using a 2D box kernel of width of 3 pixels (0.6\arcsec). The central $1\arcsec\times1\arcsec$ region used for the quasar PSF normalization 
    is masked and excluded from analysis. 
    For $\rm{BR1202\textsc{-}0725}$, we indicate the position of the \lya emitter (LAE) and SMG companions \citep[][]{Carilli2013,Drake2020} 
    with a green and white cross, respectively.}
    \label{fig:sbmaps-qso}
\end{figure}

\begin{figure}
    \centering
    \includegraphics[width=0.45\textwidth]{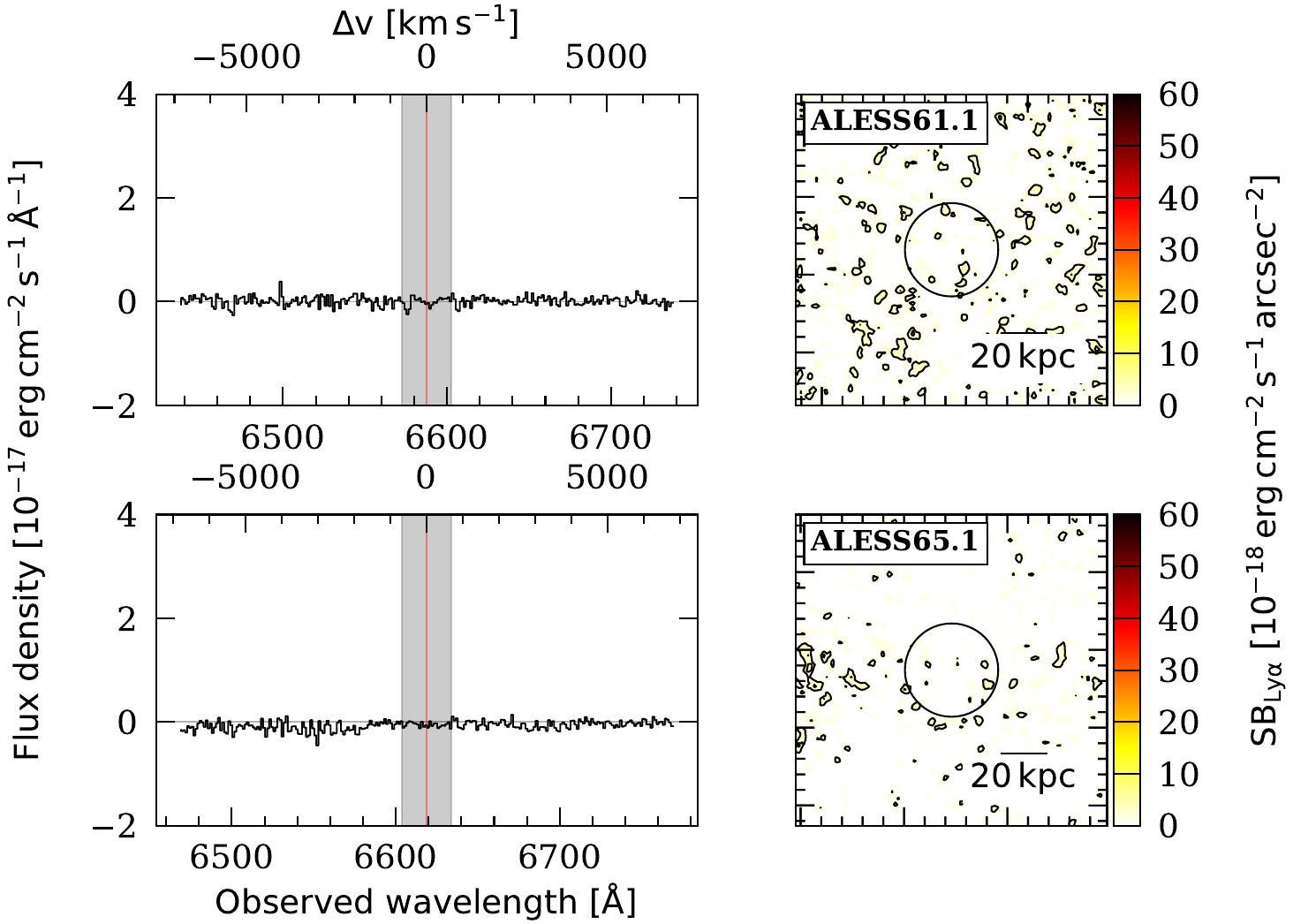}
    \caption{Integrated spectra and surface brightness maps of the two obseved SMGs. \textit{Left:} Spectra integrated inside of the $2\arcsec$ radius aperture (black circle) of ${\rm ALESS61.1}$ (top) and ${\rm ALESS65.1}$ (bottom). The gray shaded area shows the channels used to build the surface brightness map, the magenta vertical line indicates the expected \lya wavelength. \textit{Right:} \lya surface brightness maps of the two SMGs at the expected wavelength from their redshift shown in Table~\ref{tab:Data}. The maps have a side of 20\arcsec, which at their redshifts correspond to about 133 and 132\,kpc for ${\rm ALESS61.1}$ and ${\rm ALESS65.1}$, respectively. A scalebar is shown for each map. These maps are smoothed using a 2D box kernel of width of 3 pixels (0.6\arcsec). The black circles indicate an aperture of radius $2\arcsec$ centered at the SMG position.}
    \label{fig:sbmaps-smg}
\end{figure}

To ease comparison with early narrow-band studies \citep[e.g.,][]{Cantalupo2014}, here we obtain SB maps by extracting ${\rm 30\,\angstrom}$ narrow-band maps centered at the peak of the \lya extended emission from the continuum- and PSF-subtracted cubes. The peak is computed from the Gaussian fit to the integrated spectra inside a ${\rm 3\arcsec}$ radius aperture centered at the quasar position after masking the 1\,${\rm arcsec^2}$ normalization region. To test possible deviations from a Gaussian fit due to e.g. radiative transfer effects, we tested our calculation of the peak by estimating it using the first moment of the flux distribution within $\pm2.5\times$FWHM of that fit. We found agreement between the two calculations.  
The SB maps for the quasar fields in our sample are shown in the right panel of Figure~\ref{fig:sbmaps-qso}, and highlight the different morphologies and brightness that we find around these sources. Specifically, ${\rm BR1202\textsc{-}0725}$ shows the largest \lya luminosity ($\rm{\sim3\times10^{44}\,erg\,s^{-1}}$, or ${\rm \sim 2.7\times10^{44}\,erg\,s^{-1} }$ if we mask the LAE inside a $1.6\arcsec$ diameter aperture) and extent ($A=3062\,{\rm pkpc^{2}}$) (see Table~\ref{table:SFRv2}). The two obscured quasars show a factor of ${\rm \sim10}$ times smaller \lya luminosity and ${\rm \sim3}$ times smaller extent (see Table~\ref{table:SFRv2}). 
An extended \lya halo is not detected around the two SMGs at the current depth. 
This can be appreciated from the ${\rm 30\,\angstrom}$ SB maps at the expected \lya wavelength shown in the right column of Figure~\ref{fig:sbmaps-smg}, 
and also confirmed by inspecting their integrated spectra within a $2\arcsec$ radius aperture (see left column of Figure~\ref{fig:sbmaps-smg}). Appendix~\ref{sec:appendix_lines} presents the data shown in Figure~\ref{fig:sbmaps-qso} and \ref{fig:sbmaps-smg} as $\chi_{\rm smooth}$ maps (e.g., \citealt{Hennawi2013}).

Given that the detected nebulae are at a different redshifts, we test their extent also by using a common SB threshold corrected for cosmological dimming and compare to another sample presented in \citet{FAB2023}. This work report luminosity--area relations for Ly$\alpha$ nebulae around $z\sim2-3$ type-I quasars obtained using different SB thresholds. Because of the redshift of the targets and the sensitivity of our data, to compare the detected nebulae we use the reference relation, the $2\times$ and $4\times$ higher SB thresholds in that work for G09-0902+0101, G15-1444+0044, and BR1202-0725, respectively. By using these specific SB isophote, we find physical areas of $878$, $651$, $2651$\,kpc$^2$ and Ly$\alpha$ nebula luminosities of $7.5\times10^{42}$, $9.0\times10^{42}$, $18.9\times10^{43}$~erg~s$^{-1}$, respectively. These values are in agreement with the relations reported in \citet{FAB2023}, indicating that also type-I quasars hosted by SMGs sit on these luminosity-area relations. Larger samples are needed to confirm this finding.

Further, we extract an integrated spectrum for the \lya emission of each detected nebula by using all the spaxels within the ${\rm 2\sigma}$ contour of that nebula, shown in the left column of Figure~\ref{fig:sbmaps-qso}). We fit a simple Gaussian profile to describe the linewidth of the nebulae spectra. We find that all the spectra have consistent broad (${\rm FWHM\sim1500\,km\,s^{-1}}$) line profiles, with the presence of \lya absorption features. Such broad lines are not frequently seen around quasars (${\rm FWHM\sim600\,km\,s^{-1}}$; e.g., \citealt[][]{Borisova2016,FAB2019}), possibly indicating more turbulent gas reservoirs in the highly star-forming environments targeted here. Interestingly, BR1202-0725 and G09-0902+0101 show integrated \lya lines with positive shifts of the order of ${\rm 300-400\,km\,s^{-1}}$ with respect to the systemic redshift of each system, possibly indicating bulk winds.
We note that G15-1444-0044 also show a similar shift when using the maximum of its \lya line as reference velocity (see Table~\ref{table:SFRv2}). These shifts are larger than the average positive shift found for quasars with good systemic redshifts (from e.g., [CII]\,$158\ \mu$m, $69\pm36$\,km\,s$^{-1}$; \citealt{Farina2019}). We examine in detail the presence of an outflow for BR1202-0725 in Section~\ref{sec:BR1202_kinematics}. 

\begin{table*}
\centering
\caption{Summary of the properties of the detected nebulae and upper limits for non-detections.} 

\begin{tabular}{|c|c|c|c|c|c|}
 \hline
 \hline
 Object & $\rm{BR1202\textsc{-}0725}$ &  $\rm{G09\textsc{-}0902\textsc{+}0101}$ & $\rm{G15\textsc{-}1444\textsc{-}0044}$ & $\rm{ALESS61.1}$ &  $\rm{ALESS65.1}$ \\
 \hline 
 \hline
 ${\rm L_{Ly\alpha}^{QSO\ (a)}}$ & $193.6\pm0.1$ & $14.3\pm0.1$ & $13.1\pm0.1$ & -- & -- \\
 ${\rm [\times10^{43}\,erg\,s^{-1}]}$ &  &  &  &  &  \\
 \hline
 ${\rm L_{Ly\alpha}^{Neb\ (b)}}$ & $30.4\pm0.5$ & $3.4\pm0.1$ & $3.3\pm0.1$ & $<0.2$ & $<0.2$ \\
 ${\rm [\times10^{43}\,erg\,s^{-1}]}$ &  &  &  &  &  \\
 \hline

 ${\rm L_{\ion{C}{iv}}^{Neb\ (c)}}$ & $<2.3$ & $<0.2$ & $<0.1$ & $<0.7$ & $<0.8$ \\
 ${\rm [\times10^{43}\,erg\,s^{-1}]}$ &  &  &  &  &  \\
 \hline
 ${\rm L_{\ion{He}{ii}}^{Neb\ (c)}}$ & $<4.5$ & $<0.1$ & $<0.1$ & $<1.0$ & $<1.0$ \\
 ${\rm [\times10^{43}\,erg\,s^{-1}]}$ &  &  &  &  &  \\
 \hline
 
 \lya Nebula Area${\rm ^{(d)}}$ & $73.0$ ($3062$) & $20.6$ ($1195$) & $17.5$ ($961$) & -- & -- \\
 $\rm{[arcsec^2]\ (pkpc^2)}$ &  &  &  &  &  \\

 \hline
  ${\rm SFR_{Ly\alpha}^{Neb}}{\rm ^{(e)}}$ & $276.9\pm4.1$ & $30.6\pm0.9$ & $30.2\pm0.8$ & $<2$
  & $<2$\\
 ${\rm [\msun\,yr^{-1}]}$ &  &  &  &  & \\
 \hline 
 
 $\rm{\Delta v_{Neb-Sys}^{Ly\alpha}}{\rm ^{(f)}}$ & $319\pm29$ & $465\pm273$ & $39\pm272$ & -- & -- \\
 $\rm{[km\,s^{-1}]}$ &  &  &  &  & \\
 
 \hline
 
 $\rm{FWHM_{Ly\alpha}^{Neb}}{\rm ^{(g)}}$ & $1603\pm57$ & $1578\pm93$ & $1219\pm81$ & -- & -- \\
 $\rm{[km\,s^{-1}]}$ &  &  &  &  &  \\
 
 \hline

 ${ \rm SB_{ \ion{Si}{iv} } }$ & $<0.5$ & $<0.6$ & $<0.7$ & $<0.7$ & $<1.2$ \\
 ${ \rm \times[10^{-18}]^{(h)} }$ &  &  &  &  & \\
 \hline
 
 ${ \rm SB_{ \ion{C}{iv} } }$ & $<1.0$ & $<0.8$ & $<0.6$ & $<1.3$ & $<1.3$ \\
 ${ \rm \times[10^{-18}]^{(h)} }$ &  &  &  &  & \\
 \hline
 
 ${ \rm SB_{ \ion{He}{ii} } }$ & $<1.9$ & $<0.5$ & $<0.6$ & $<1.8$ & $<1.7$\\
 ${ \rm \times[10^{-18}]^{(h)} }$ &  &  &  &  & \\
 \hline
 
 ${ \rm SB_{ \ion{C}{iii]} } }$ & -- & $<1.2$ & $<1.8$ & -- & -- \\
 ${ \rm \times[10^{-18}]^{(h)} }$ &  &  &  &  & \\
 \hline

\end{tabular}

\flushleft 
${ \rm ^{(a)} }$: Quasars' Ly$\alpha$ luminosities obtained by integrating each quasar spectrum in $\pm$FWHM of the nebular emission.

${\rm ^{(b)}}$: Ly$\alpha$ luminosities within the $2\sigma$ contours of Figure~\ref{fig:sbmaps-qso} for the three quasars. For the two SMGs, $2\sigma$ upper limits rescaled assuming a nebula area of $20\,{\rm arcsec}^2$ (from the 30\,\AA\ narrow-band).

${\rm ^{(c)}}$: $2\sigma$ upper limits on the \ion{C}{iv} and \ion{He}{ii} luminosities. For the three quasars, these are obtained for the \lya nebula area within the the same velocity range as the fitted \lya line ($\rm{\pm FWHM_{Ly\alpha}^{Neb}}$). For BR1202-0725, most of the velocity range redwards of the 
\ion{He}{ii} line falls outside of the spectral range of the MUSE datacube, therefore we use the same velocity span but shifted to the last available channels. 
While for the SMGs, the limits are obtained assuming a nebula area of $20\,{\rm arcsec}^2$ within a velocity range equivalent to the \lya 30\,\AA\ narrow-band.

${\rm ^{(d)}}$: Area of the Ly$\alpha$ nebulae inside the $2\sigma$ contour.

${\rm ^{(e)}}$: SFR obtained from the Ly$\alpha$ luminosities listed in this table, assuming case B recombination and Formula 2 in \citet{Kennicutt1998} (see section~\ref{subsec:starformpower}).

${\rm ^{(f)}}$: Velocity shift between the quasars' systemic redshift and the Ly$\alpha$ peak velocity of the integrated nebula spectrum computed as first moment. For completeness, the velocity shifts computed using simply the maximum of each spectrum are $332\pm29$, $444\pm273$, and $382\pm272$~km~s$^{-1}$ for BR1202-0725, G09-0902+0101, and G15-1444-0044, respectively.

${\rm ^{(g)}}$: FWHM of the Gaussian fit of the integrated Ly$\alpha$ emission shown in Figure~\ref{fig:sbmaps-qso}.

${\rm ^{(h)}}$: $2\sigma$ surface brightness limits for an aperture equivalent to the nebula area (assuming 20 arcsec$^2$ for the ALESS SMGs), and in a velocity range equivalent to the $30\,\angstrom$ narrow-band used for Ly$\alpha$, in units of ${ \rm erg\,s^{-1}\,cm^{-2}\,arcsec^{-2} }$. 

\label{table:SFRv2}
\end{table*}

\subsection[Ly\texorpdfstring{$\alpha$}{alpha} surface brightness radial profiles]{Ly$ \alpha$ surface brightness radial profiles}\label{subsec:radialprofiles}

\begin{figure}
    \centering
    \includegraphics[width=0.45\textwidth]{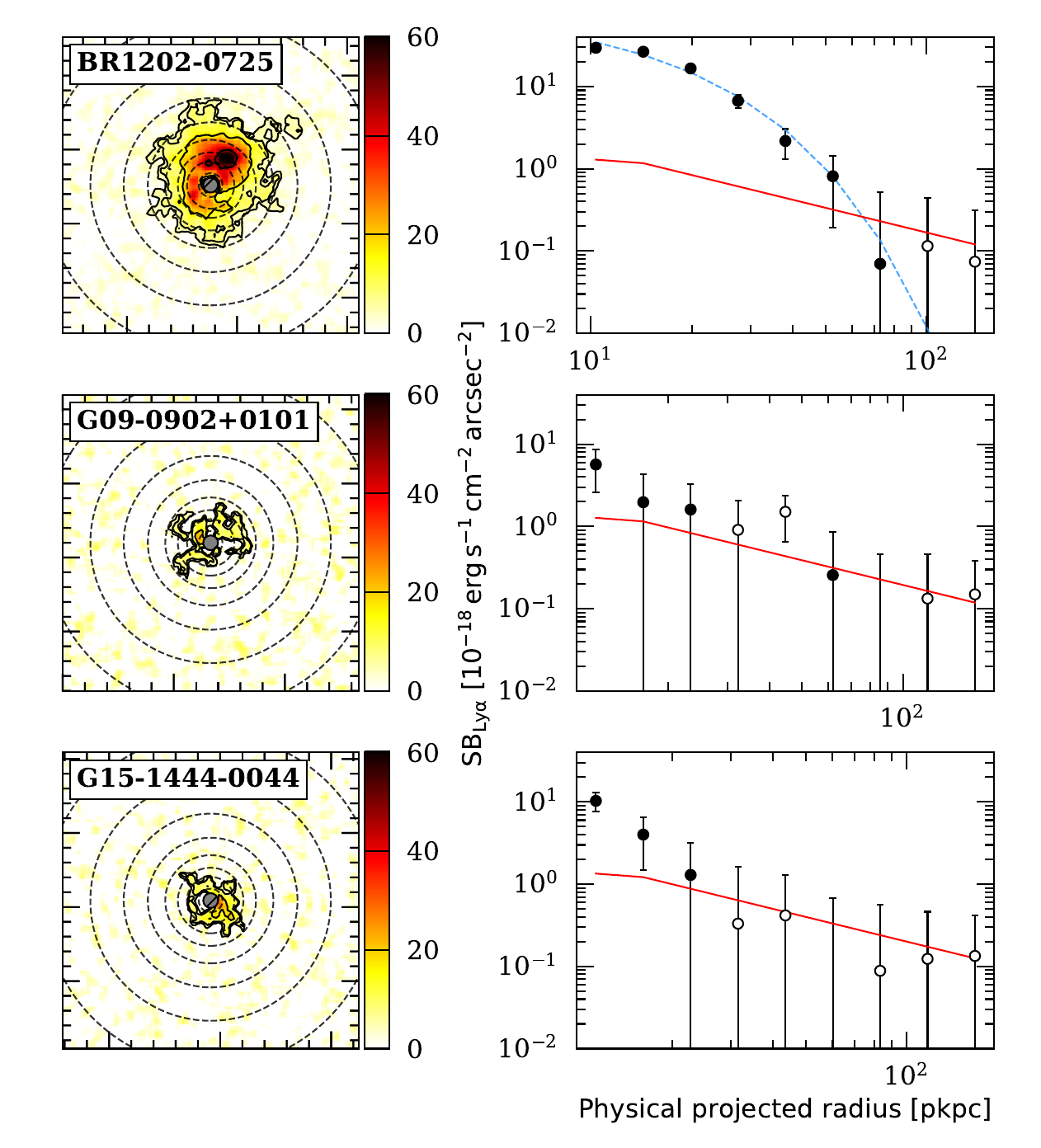}
    \caption{\lya surface brightness radial profiles extraction. \textit{Left:} Smoothed SB maps of the quasar sample (same as in Figure~\ref{fig:sbmaps-qso}) with the circular apertures used to extract radial profiles overlayed.
    The gray circle in each map indicates the 1~arcsec$^2$ region used for the quasar's PSF normalization, which is neglected in our analysis. \textit{Right:} Extracted radial profiles inside each annulus. Open circles indicate negative values shown here for completeness. The red line represents the $2\sigma$ SB limit rescaled to the area of each annulus. The blue dashed curve in the top panel is the best fit of the datapoints above the red line using an exponential function with the form ${\rm SB}_{Ly\alpha}=C\exp{(-r/r_h)}$, where $C=85.3\pm13.8\times10^{-18}\,{\rm erg\,s^{-1}\,cm^{-2}\,arcsec^{-2}}$ and $r_h=11\pm1$\,pkpc.}
    \label{fig:maprofiles}
\end{figure}

In order to describe the morphology and relative brightness of our nebulae, we build circularly averaged radial profiles as usually done in the literature \citep[e.g.,][]{FAB2019,Fossati2021}. The radial profiles are extracted from the ${\rm 30\,\angstrom}$ SB maps from Figures~\ref{fig:sbmaps-qso}, using logarithmically spaced bins starting at about ${\rm 1.6\arcsec}$. The left panels of Figure~\ref{fig:maprofiles} show the circular apertures used to extract the profiles overlayed on the SB maps for the detected sources, while the right panels show the individual extracted profiles. Filled circles represent positive mean SB values, while open circles represent negative mean SB values. In the same figure, the solid red line represents the ${\rm 2\sigma}$ SB limit within each ring. From these profiles it is evident that the QSO+SMG system
BR1202-0725 has a more extended and bright nebula compared to the two SMGs hosting a QSO.
We fit the observed SB profile of BR1202-0725 with both an exponential ${\rm SB}_{\rm Ly\alpha}=C\exp{(-r/r_{\rm h})}$ and a power law ${\rm SB}_{\rm Ly\alpha}=Cr^{\alpha}$ as usually done in the literature, finding a better agreement with an exponential function. The scale length for the best fit is $r_h=11\pm1\,$kpc.  We do not perform any fit on the other two detected nebulae given the limited radial range covered. We discuss further these SB profiles in comparison to other samples in the literature in Section~\ref{subsec:SBprofilesDiscussion}.

%%%%%%%%%%%%%%%%%%%%%%%%%
%       KINEMATICS      %
%%%%%%%%%%%%%%%%%%%%%%%%%

\subsection[Ly\texorpdfstring{$\alpha$}{alpha}]{Ly$ \alpha$ kinematics}\label{subsec:kinematics}

\subsubsection{Mean velocity shift and velocity dispersion maps}\label{subsec:meanshift_sigma}

\begin{figure}
    \centering
    \includegraphics[width=0.45\textwidth]{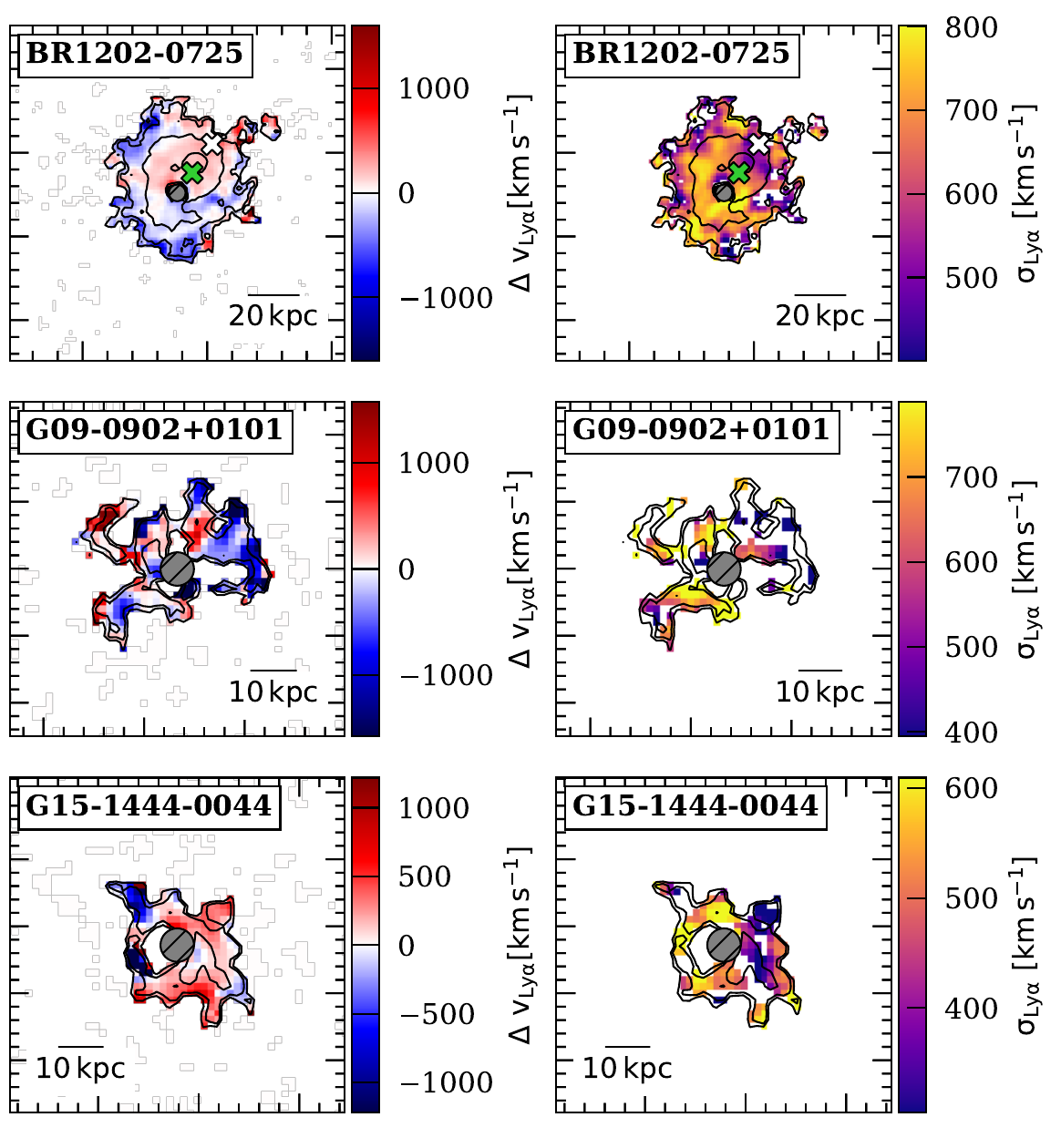}
    \caption{Kinematics of the detected \lya nebulae. Velocity shift $\Delta v_{\rm Ly\alpha}$ maps (left column) with respect to the \lya peak wavelength of the nebula (see Figure~\ref{fig:sbmaps-qso}), and velocity dispersion $\sigma_{\rm Ly\alpha}$ maps (right column) of the \lya nebula around the three quasars, computed from the first and second moment, respectively. The maps are smoothed as in Figure~\ref{fig:sbmaps-qso}. 
    The moment calculation is performed only inside the $2\sigma$ isophote of each nebula. The position of the LAE and SMG companion of the BR1202-0725 system are marked with a green and white cross, respectively. The side of the maps is 20\arcsec for BR1202-0725, and 10\arcsec for G09-0902+0101 and G15-1444-0044. A scalebar is shown for each map. The gray circle in each map indicates the 1~arcsec$^2$ region used for the quasar's PSF normalization, which is neglected in our analysis.}
    \label{fig:kinematic-maps}
\end{figure}

We compute velocity shift and velocity dispersion maps for our detected sources by building the first and second moment maps within narrow wavelength ranges, which are centered at the peak of their nebula \lya spectrum and encompass $\pm$FWHM 
of their gaussian fit (see Table~\ref{table:SFRv2}). Figure~\ref{fig:kinematic-maps} shows the velocity shift maps and the velocity dispersion maps on the left and right columns respectively. The velocity shift maps are computed with respect to the \lya line peak of each nebula and highlight the complex kinematics of the nebulae, likely affected by some \lya resonant scattering. We select the \lya line peak as reference to keep our sample homogeneous as we do not have an accurate systemic redshift for the SMGs hosting QSOs. We find that $\rm{BR1202\textsc{-}0725}$ presents a clear blueshifted and redshifted component at velocities of $\sim\pm 400$\,km\,s$^{-1}$. Larger velocity dispersions are seen close to the quasar (FWHM$\sim 1600-1800$\,km\,s$^{-1}$), while we observe that the companion SMG (indicated with a white cross) is located, in projection, in a region with more quiescent kinematics with respect to the rest of the nebula (FWHM$\sim1000$\,km\,s$^{-1}$). This might be due to the fact that this zone is less turbulent or that dust absorbs some of the \lya photons 
\citep[e.g.,][]{Laursen2009}. However, we stress that there is a substantial velocity shift between the SMG ($-142.1\,{\rm km\,s^{-1},}$ with respect to the systemic $z$; \citealt{Drake2020}) and the Ly$\alpha$ emission at its projected location ($\sim +400$~km~s$^{-1}$), possibly indicating the presence of peculiar velocities between the galaxy and the gas, on top of radiative transfer effects. Similarly, LAE1 shows a velocity offset of $+42$~km~s$^{-1}$ with respect to BR1202-0725 (\citealt{Drake2020}), which is smaller than the \lya velocity offset at its projected location. 
We further study the large velocity components near the quasar in the BR1202-0725 system in Section~\ref{sec:BR1202_kinematics}.

The velocity shift and dispersion maps 
for $\rm{G09\textsc{-}0902\textsc{+}0101}$ and $\rm{G15\textsc{-}1444\textsc{-}0044}$ appear noisy, possibly due to  
their lower surface brightness levels and the complexity of the gas motions on these scales (\citealp[e.g., Figure~B1 in][]{Costa2022}). Deeper observations would be needed to study in detail their kinematics. 

\subsubsection{Evidence of violent kinematics in BR1202-0725}\label{sec:BR1202_kinematics}

\begin{figure*}
    \centering
    \includegraphics[width=0.8\textwidth]{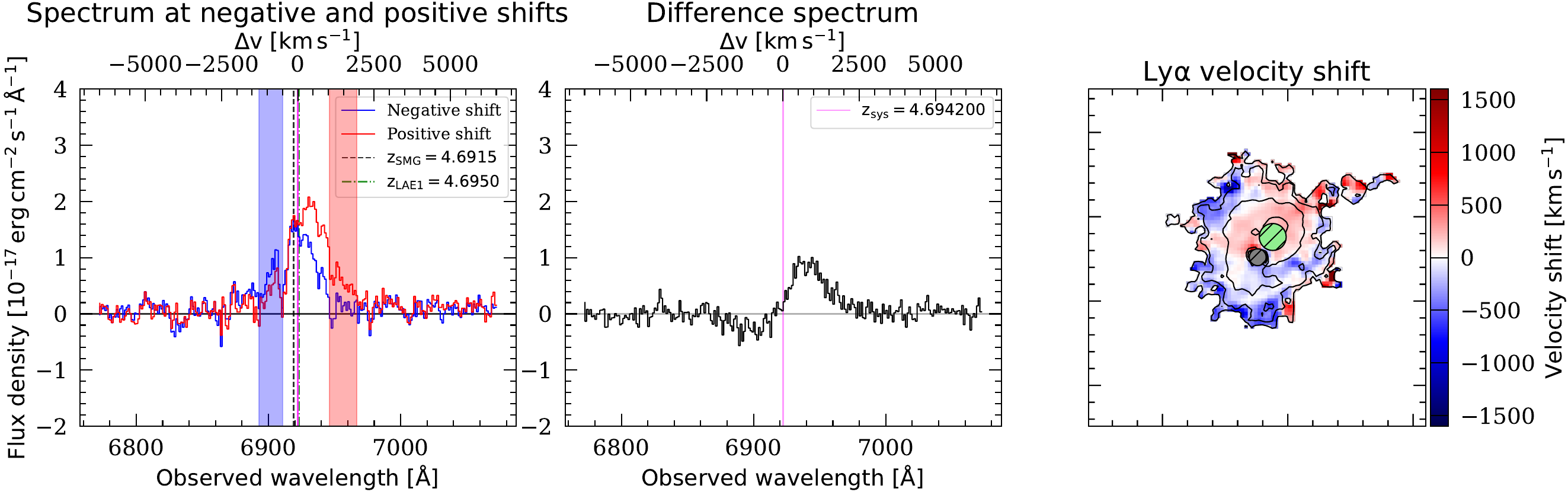}\\
    \vspace{0.3cm}
    \includegraphics[width=0.57\textwidth]{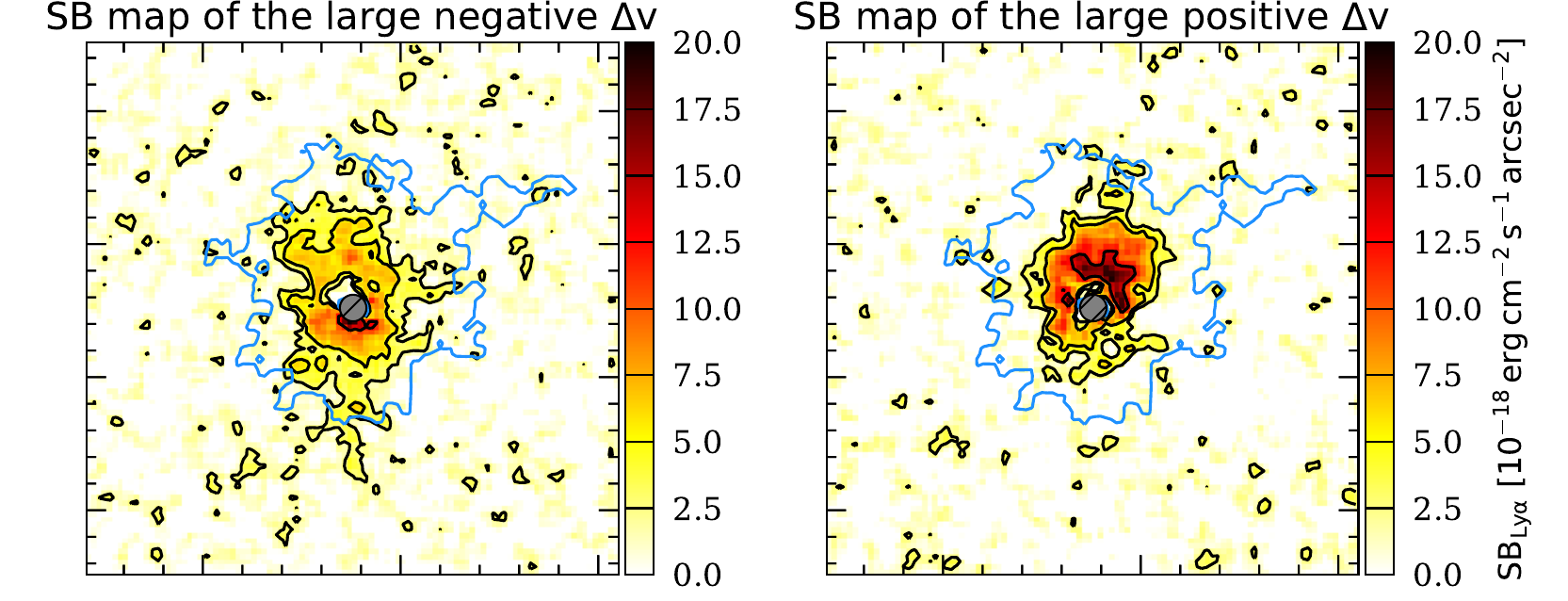}
    \caption{Violent kinematics in the \lya nebula of BR1202-0725. {\it Top left:} Integrated spectra of regions with negative (blue curve) and positive (red curve) velocity shifts for BR1202-0725, with respect to the peak of the nebular \lya emission (see Figure~\ref{fig:sbmaps-qso}). The spectra are taken after masking the $1\arcsec\times1\arcsec$ region of the quasar's PSF subtraction and a circular region with diameter  
    of 1.6\arcsec at the LAE companion position. We indicate with vertical lines the systemic redshift (solid magenta), the redshift of the SMG (black dashed) and LAE (green dot-dashed) companions \citep[][]{Carilli2013,Drake2020}. 
    {\it Top middle:} Residual spectrum between the positive and negative shift spectrum. We indicate the systemic redshift with a magenta line. 
    {\it Top right:} \lya velocity shift map computed from the first moment using a spectral window of $\pm$FWHM centered at the peak of the nebular \lya emission (see Table~\ref{table:SFRv2}) and smoothed as in Figure~\ref{fig:sbmaps-qso}. The position of the QSO and LAE companion masks are indicated with a gray and green hashed circle, respectively.  {\it Bottom left:} \lya SB map using the spectral channels with the largest negative shifts of the nebulae, indicated with a shaded blue region in the top left panel. The blue contours indicate the $2\sigma$ isophote of the extended emission shown in 
    Figure~\ref{fig:sbmaps-qso}. {\it Bottom right:} \lya SB map using the spectral channels with the largest positive shifts of the nebulae, indicated with a shaded red region in the top left panel. Both of the SB maps at the bottom are smoothed as in Figure~\ref{fig:sbmaps-qso}. All maps have a side of $20\arcsec$ and black contours indicating surface brightness levels of $[2, 4, 10, 20, 50]$~$\sigma$ within the respective wavelength ranges.} 
    \label{fig:BR1202_outflow}
\end{figure*}

As mentioned in Section~\ref{subsec:meanshift_sigma}, we observe large \lya velocity dispersions near the QSO position in the BR1202-0725 system. Indeed, we estimate a large \lya FWHM (see Table~\ref{table:SFRv2}), which could hint to the presence of large-scale outflows due to the QSO and star-formation activity. Similarly large \lya FWHM ($>1000$~km~s$^{-1}$) has been observed in the CGM around a broad-absorption line (BAL) quasar \citep[][]{Ginolfi2018}, suggesting that we observe broader \lya emission when there are outflow signatures on small scales. 

In this section, we further explore the aforementioned violent kinematics found in the \lya nebular emission of BR1202-0725. This nebula is bright and therefore we can access the large velocity wings of its Ly$\alpha$ line. Specifically, we investigate the difference in the Ly$\alpha$ line shape in the regions with positive and negative velocity shifts as seen from its first moment map (see Figure~\ref{fig:kinematic-maps}). We show in the top left panel of Figure~\ref{fig:BR1202_outflow} the integrated spectra for regions with negative (blue) and positive (red) velocity shifts, with respect to the peak of the nebular \lya emission.
 For the integrated spectrum at positive velocities we subtract the emission coming from a circular region of radius 1.6\arcsec centered at the LAE companion position (from \citealt{Drake2020}; see top-right panel in Figure~\ref{fig:BR1202_outflow}) to be sensitive only to the emission from large-scale gas.
Assuming this LAE follows a rescaled version of the average SB profile for LAE presented by \citet{Wisotzki2018}, extended emission associated with the LAE outside the chosen aperture should be below our $2\sigma$ detection threshold, and therefore negligible.

We observe that both these spectra display large linewidths with wings at large velocities ($\Delta v>1000$\,km\,s$^{-1}$). However, the spectrum for the positive velocities shows an excess with a wing out to velocities as high as $\sim2500$\,km\,s$^{-1}$ that can be appreciated in the difference spectrum reported in the middle panel of Figure~\ref{fig:BR1202_outflow}.
Also, both spectra show the presence of an absorption feature at velocities $-470$\,km\,s$^{-1}$. This feature was also reported in \citet{Drake2020}, who suggested that it is due to a shell of outflowing material as usually invoked for the case of HzRGs (e.g., \citealt{vanOjik1997}) and quasar pairs (e.g., \citealt{Cai2018,FAB2019b}). 

Further, we investigate from which spatial region the largest velocity shifts are observable. The bottom panels of Figure~\ref{fig:BR1202_outflow} show \lya SB maps of BR1202-0725 using velocity channels corresponding to the largest velocity shifts, indicated with blue and red shaded regions in the spectra at the top panel respectively. These maps show that the largest velocity shifts are located closer to the quasar position, 
extending out to about $20\,$kpc at the current depth. While all the described features are reminiscent of an outflow, radiative transfer modelling is needed to test this scenario (e.g., \citealt{Chang2023}).  
Such a modelling is ongoing and will be presented in a future publication. 

\subsection{Searching for \ion{C}{iv} and \ion{He}{ii} extended emission}
\label{sec:otherLines}

As mentioned in the introduction, the emission lines \ion{C}{iv} and \ion{He}{ii} could be used to 
constrain the powering mechanism responsible for the observed nebulae, the hardness of the ionizing spectrum of the powering source and shock scenarios, the metallicity, and the extent of the enriched halo.
For this reason we search for these lines in the final datacubes, 
together with additional rest-frame UV emission lines, \ion{Si}{iv} and \ion{C}{iii}], that could be used to constrain the physical properties of the extended gas.
For all the sources in our sample, we do not detect extended emission in these transitions. For completeness, we present in Figures~\ref{fig:linesBR1202},~\ref{fig:linesG09},~\ref{fig:linesG15},~\ref{fig:linesA61}~and~\ref{fig:linesA65} of Appendix~\ref{sec:appendix_lines} $\chi_{\rm Smooth}$  maps obtained for velocity ranges equivalent to the \lya 30\,\AA\ NB  
and centered at the expected wavelengths for \ion{Si}{iv}, \ion{C}{iv}, \ion{He}{ii} and \ion{C}{iii]}, respectively. For the BR1202-0725 system, the \ion{He}{ii} line is within the datacube spectral range, but the equivalent NB falls outside of the spectral range. Therefore we use the last available channels to compute this $\chi_{\rm Smooth}$ map. 
Since there are no detections for any of these additional lines, we compute $2\sigma$ SB upper limits and rescale them to their respective \lya nebula area and assuming 20 squared arcseconds for the ALESS SMGs, these values are shown in Table~\ref{table:SFRv2}.

We estimate the \ion{C}{iv}/\lya and \ion{He}{ii}/\lya ratio upper limits for the three systems with QSOs and discuss them in the context of the \lya powering mechanisms in Section~\ref{sec:poweringmechanisms}. These are estimated from the total \lya luminosity of each nebula and the luminosity upper limits of \ion{C}{iv} and \ion{He}{ii} (shown in Table~\ref{table:SFRv2}). For BR1202-0725, the \ion{He}{ii} velocity range equivalent to the $\pm$FWHM of the \lya line is outside of the spectral range, therefore we use the same velocity span encompassing the line, but shifted to the last available channels to estimate this limit. Therefore, we caution that this upper limit may not fully represent the noise in the range $\pm$FWHM of the \ion{He}{ii} line. However, as the \ion{He}{ii} line is expected to be narrower than the \lya emission, the range used surely quantifies the absence of a \ion{He}{ii} detection in the current observation.

%%%%%%%%%%%%%%%%%%
%   DISCUSSION   %
%%%%%%%%%%%%%%%%%%

\section{Discussion}\label{sec:discussion}

\subsection{Extended Lyman-$\alpha$ powering mechanisms}\label{sec:poweringmechanisms}
The variety of the systems studied here (QSO+SMG, two SMGs hosting a QSO, two SMGs) enables us to discuss the different mechanisms through which extended \lya nebular emission could be produced. 
We will consider the following mechanisms: scattering of \lya photons produced in compact sources, ionizing and \lya photons produced by ongoing star formation, gravitational cooling radiation, shocks by galactic and/or AGN outflows, and AGN photoionization followed by recombination. In principle, all these mechanisms could contribute together to the production of the observed extended emission and we discuss their relative roles below. 
In brief, by revisiting
each of the aforementioned mechanisms 
we show that the firmest conclusion is that gravitational cooling is likely not contributing significantly in powering the observed extended \lya emission.

\subsubsection[Ly\texorpdfstring{$\alpha$}{alpha}]{Scattering of Ly$\alpha$ photons emitted by compact sources}\label{subsec:scatteringpower}
\lya photons produced in compact sources can scatter out in the surrounding gas distribution, producing an observable glow.
In this scenario, the surface brightness of the extended nebular \lya emission is therefore expected to be proportional to the \lya photon budget from the QSO \citep[e.g.,][]{Hennawi2013} and/or embedded galaxies, and 
no significant extended \ion{He}{ii} emission should be detected, as it is a recombination line \citep[e.g.,][]{Prescott2015, FAB2015a, FAB2015b}.

For the systems studied here, the dominant contribution of \lya photons in BR1202-0725, G09-0902+0101 and G15-1444-0044 
is clearly the AGN. Indeed, in close proximity to the extended \lya emission BR1202-0725 has three LAEs ($F_{\rm LAE1} =(1.54 \pm 0.05)\times 10^{-16}\, {\rm erg\,s^{-1}\,cm^{-2}}$, $F_{\rm LAE2} =(0.54 \pm 0.05)\times 10^{-16}\, {\rm erg\,s^{-1}\,cm^{-2}}$ and $F_{\rm LAE3} =(0.24 \pm 0.03)\times 10^{-16}\, {\rm erg\,s^{-1}\,cm^{-2}}$; \citealt{Drake2020}), but their \lya emission is only $3$\% of the quasar budget at those wavelengths (see Table~\ref{table:SFRv2}; $(8.661\pm 0.004) \times 10^{-15}\,{\rm erg\,s^{-1}\,cm^{-2}}$). Instead, G09-0902+0101 and G15-1444-0044 do not show LAEs in close proximity to their extended \lya glow, while in the SMG fields we do not find evidence for any \lya emitting sources. 
We compute an indicative number of \lya photons available for scattering by integrating the quasar's spectra using the $\pm$FWHM of the nebular emission (luminosities listed in Table~\ref{table:SFRv2}).
We find that for each quasar system these photons would be able to power the observed \lya nebulae, with the quasar \lya luminosities exceeding by 6.4, 4.3 and 4.0 the nebular luminosities (computed in the same range; Table~\ref{table:SFRv2}) for $\rm{BR1202\textsc{-}0725}$, $\rm{G09\textsc{-}0902\textsc{+}0101}$ and $\rm{G15\textsc{-}1444\textsc{-}0044}$, respectively. 

Interestingly, the \lya luminosity of the quasar in the BR1202-0725 system is $14\times$ greater than $\rm{G09\textsc{-}0902\textsc{+}0101}$ and $\rm{G15\textsc{-}1444\textsc{-}0044}$. In this scenario, if the conditions and environment of these three systems are similar, this would predict a $14\times$ more luminous nebula around BR1202-0725. However, the dust mass in BR1202-0725 is on average $1.3\times$ greater than in the two SMGs hosting QSOs, likely corresponding to a larger probability of \lya absorption. Therefore, we would only expect a $10\times$ more luminous nebula. This calculation aligns well with the observation that the nebula around BR1202-0725 is $9\times$ more luminous than $\rm{G09\textsc{-}0902\textsc{+}0101}$ and $\rm{G15\textsc{-}1444\textsc{-}0044}$.
Accordingly, the two SMGs, where no strong source of \lya photons is observed, do not have an extended nebula. Therefore the systems studied here are consistent with a picture in which the extended \lya level could be regulated by a combination between the QSO's budget of \lya photons and the dust content of the host galaxy.

\begin{figure}
    \centering
    \includegraphics[width=\columnwidth]{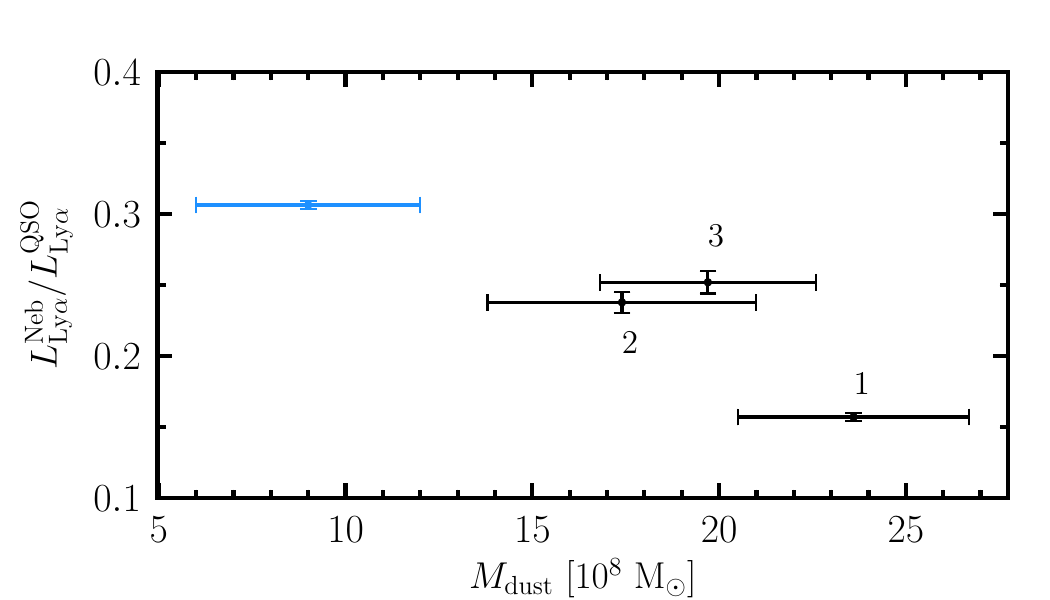}
    \caption{Ratio of the nebula and quasar Ly$\alpha$ luminosities versus the dust mass of the central host galaxy for the three quasars in our sample (black with IDs) and for the Fabulous system (blue; \citealt{FAB2022}).}
    \label{fig:new}
\end{figure}

To further test this trend, we obtain the nebula and quasar Ly$\alpha$ luminosities for another system with known extended emission and dust mass estimate ($M_{\rm dust}=(9\pm3)\times10^8$~M$_{\odot}$; \citealt{FAB2022}), finding $L_{\rm Ly\alpha}^{\rm Neb}=(33.7\pm0.3)\times10^{43}$\,erg\,s$^{-1}$ and $L_{\rm Ly\alpha}^{\rm QSO}= (110.0\pm0.1)\times10^{43}$\,erg\,s$^{-1}$. We find that also this system follows the aforementioned tentative trend. To visualize this finding, Figure~\ref{fig:new} shows how the ratio $L_{\rm Ly\alpha}^{\rm Neb}/L_{\rm Ly\alpha}^{\rm QSO}$ decreases with $M_{\rm dust}$.

We stress that previous works looking at variation of \lya luminosities for quasars' nebulae against the QSO's budget of \lya photons and/or the QSO's luminosity \citep[e.g.,][]{FAB2019,Mackenzie2020} found that if a relation is present, its scatter is large. However, those studies did not have any information on the dust properties of the host galaxies, which could be 
an important factor  
in finding a  
trend.  
Larger samples of QSO \lya nebulae with well-studied QSO hosts are required to assess the contribution from this scenario.

Moreover, the detected nebulae present an anisotropic distribution and an offset between the quasar position and the peak of the \lya emission. For example, BR1202-0725 shows no \lya emission at the position of the dusty SMG (see white cross in Figure~\ref{fig:sbmaps-qso}) and the peak of the \lya distribution is closer to the LAE (see Figure~\ref{fig:sbmaps-qso}) where we expect less dust content. The morphological differences and brightest part of the nebulae could therefore be due to the distribution of scattering atoms and dust around each QSO, with the denser regions and the dustier regions showing higher and lower \lya emission, respectively. 
Therefore, due to all this, we cannot rule out that resonant scattering of the QSO's \lya photons can contribute in powering the detected nebulae.

Finally, we stress two important caveats. First, the same morphological and surface brightness differences observed between the detected nebulae could also be due to variations in the gas physical properties (e.g., density, local dust content) in a photoionization scenario (see Section~\ref{subsec:photoionizationpower}). Secondly, differences in surface brightness could also be produced due to the different local environment in which the extended emission is observed (e.g., presence of companion galaxies) as it is the case for BR1202-0725. We discuss this possibility in Section~\ref{subsec:SBprofilesDiscussion}.

\subsubsection{Star formation}\label{subsec:starformpower}

The \lya emission could also be powered by recombination radiation which follows photoionization by the strong star formation of the targeted sources.  As any Ly$\alpha$ emission mechanism this process can be followed by resonant scattering, which we do not take into account in this section. In this scenario, the \lya luminosity is expected to be proportional to the SFR of the embedded sources, modulo the escape fraction of their ionizing and \lya photons. We can get a rough estimate of the SFR needed to power the observed extended nebulae by assuming that all of the \lya luminosity is produced by star formation under case B recombination ($L_{\rm Ly\alpha}=8.7L_{\rm H\alpha}$) and using the Equation~2 of \citealt[][]{Kennicutt1998}: ${\rm SFR(\msun\,yr^{-1})}=L_{\rm H\alpha} /(1.26\times10^{41}\,{ \rm erg\, s^{-1} } )$. For the SMGs we do not find any extended \lya emission, therefore we report upper limits for their \lya SFRs. We compute them from their $2\sigma$ SB limit (for a 30\,\AA\ narrow band) scaled to a similar nebula area as for the two SMGs hosting a QSO (20 arcsec$^2$). All the determined SFR values are listed in Table~\ref{table:SFRv2}.  

We compare these derived instantaneous SFRs (timescales up to $\sim10$\,Myr; \citealt{KennicuttEvans2012}) with the SFRs estimated from the IR (timescales up to $\sim100$\,Myr; \citealt{KennicuttEvans2012}), assuming a constant star-formation history. 
We find that the available star-formation is larger than that estimated from the Ly$\alpha$, implying that it would be enough to power the detected Ly$\alpha$ nebulae, also when a very small ($2-5\%$) escape of ionizing photons is taken into account.
Interestingly, a similar result was found for $z\sim6$ quasars (median value $\lesssim1$\%; \citealt{Farina2019}).
However, the fact that we do not detect emission in the surrounding and proximity of the three SMGs in our sample (the two ALESS sources and the SMG companion of $\rm{BR1202\textsc{-}0725}$), run counter to the star formation scenario. Indeed, the ratios between the detected nebular emission for the quasars in our sample (which are all hosted by SMG-like galaxies) and the upper limits for SMGs are $>16-160$, much larger than the ratios between the SFRs for the targeted objects ($4-13$). In other words, if star-formation is the main powering mechanism of extended Ly$\alpha$ the SMGs should have nebulae with $L_{\rm Ly\alpha}=4.4\times 10^{43}$\,erg\,s$^{-1}$, which would be detectable in the current dataset if they spread out on scales similar to the nebular emission around the two SMGs hosting a QSO. The fact that we do not detect such emission means that most of the ultraviolet photons produced by star-formation ($>95$\%, using the 2$\sigma$ limit) do not escape these galaxies. 
We stress that the dust contents estimated for the SMGs are the smallest in our sample (Table~\ref{tab:Data}). 
Therefore it would be required a drastic change in dust geometry and dust grain properties between the different sources to ascribe the diversity of Ly$\alpha$ nebulae in this sample to star-formation.
Thus, star formation seems to have a minor role in powering the extended emission in these sources.

\subsubsection{Gravitational cooling radiation}\label{subsec:coolingpower}

Large-scale Ly$\alpha$ emission could be produced by gravitational cooling radiation, originating from cold streams accretion. \citet[][]{Rosdahl&Blaizot2012} and \citet[][]{Trebitsch2016} have done radiative transfer calculations to test this scenario, and found that it can reproduce the sizes and luminosities of extended \lya emission in massive halos, together with their polarization pattern.

In this scenario, for the high DM halo masses inferred for our sources (see Section~\ref{sec:data}), uncertain analytical and numerical calculations predict  
the extended \lya luminosities to be proportional to the DM halo masses, and the peak of the signal to be located at the center of the gravitational potential well \citep[][]{Dijkstra2009,Faucher-Giguere2010}.
We stress upfront that the DM halo masses currently available are highly uncertain (see caveats in Section~\ref{subsec:physProp}). However, it is of interest to check whether they favor a gravitational cooling scenario, also to make predictions for future observations.

For the inferred DM halo masses we would expect G09-0902+0101 to have the brightest Ly$\alpha$ glow, followed by BR1202-0725 ($1.6\times$ dimmer), and then G15-1444-0044 
and the SMGs ($10\times$ dimmer). 
However, the \lya nebulae around SMGs hosting a QSO are 9 times fainter than the nebula around the QSO+SMG system. Moreover, in this scenario the SMGs should have extended emission only two times dimmer than BR1202-0725, but they are not detected at the current depth.  
Therefore there is no clear evidence that gravitational cooling powers these nebulae. This is also supported by recent cosmological simulations of high-redshift QSO halos post-processed with radiative transfer calculations \citep[][]{Costa2022}. 
the contribution from gravitational cooling is about one order of magnitude lower compared to the effects of recombination and scattering of \lya photons from the BLRs of quasars. A factor of ten lower emission with respect to the SMGs hosting a QSO cannot be detected in the current dataset.

To search for the gravitational cooling signal one should therefore target obscured isolated galaxies, such as the SMGs studied here. Indeed, in these objects there should be only a very minor contribution from the other mentioned mechanisms.
To search for the gravitational cooling signal at a factor of ten lower SB level than for the SMGs hosting QSO studied here, one would need to be able to detect an average signal of the order of ${\rm SB}_{\rm Ly\alpha}\sim 4\times10^{-19}$\,erg\,s$^{-1}$\,cm$^{-2}$\,arcsec$^{-2}$. This depth could be achieved at high significance ($3\sigma$) with a MUSE 75~hours exposure of an individual system or with a stack of multiple systems with an equivalent total exposure time.
The detection of this signal would give an independent indication of the halo mass of SMGs and of its cool gas mass fraction.

\subsubsection{Shocks by Galactic/AGN outflows}\label{subsec:shockpower}

\begin{figure*}
    \centering
    \includegraphics[width=\textwidth]{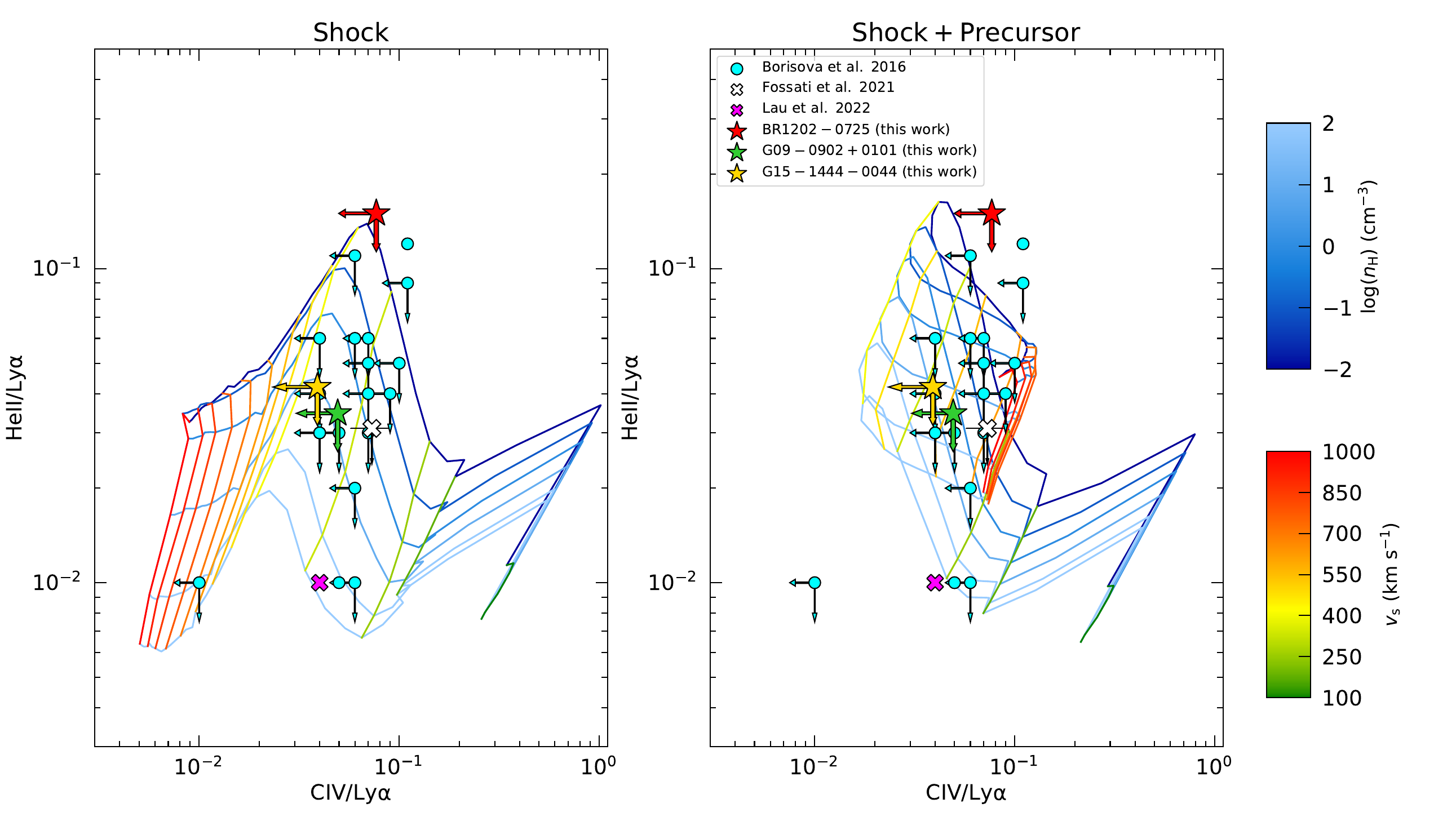}
    \caption{\ion{C}{iv}/Ly$\alpha$ and \ion{He}{ii}/Ly$\alpha$ line ratio upper limits ($2\sigma$) for the sources studied here with detected extended Ly$\alpha$ nebulae (star symbols). The line ratios are computed from the nebular luminosities listed in Table~\ref{table:SFRv2}.
    The left panel shows the model grids for the shock scenario, while the right panel shows the model grids for the shock+precursor scenario from \citet{Allen2008}. Overlayed are derived line ratios and upper limits for 
    QSO nebulae from \citet{Borisova2016} as cyan circles, stacked nebulae around $z=3-4.5$ QSOs inside a radial bin of $10<R/{\rm kpc} < 30$ from \citet{Fossati2021} as a white cross, and one extremely red quasar at $z\sim2.3$ (\citealt{Lau2022}) as a magenta cross.}
    \label{fig:line_ratios}
\end{figure*}

We observe broad \lya linewidths ($\rm{FWHM}\sim1500\,\kms$, see Table~\ref{table:SFRv2}) in all our nebulae around QSOs. These FWHM are broader than usually reported in the literature ($\rm{FWHM}< 940\,\kms$, e.g., \citealt[][]{Borisova2016,FAB2018,Fossati2021}), suggesting 
the presence of violent kinematics (shocks)
driven by the QSO and/or star-formation activity. Indeed, the sound speed in the ambient hot medium ($T\sim5\times10^6\,{\rm K}$) for a $10^{12.5}\,{\rm M_{\odot}}$ dark-matter halo is $c_{\rm s}\approx260\,{\rm km\,s^{-1}}$. 
Moreover, we observe a positive velocity shift in all our nebulae with respect to the QSO's systemic redshift (Figure~\ref{fig:sbmaps-qso}), which could further indicate the presence of bulk outflows as traced by \lya resonant scattering effects.

In order to build intuition about a shock scenario, we compare \ion{C}{iv}/Ly$\alpha$ and \ion{He}{ii}/Ly$\alpha$ line ratios to those predicted by shock and shock+precursor models from \citet{Allen2008}, as done in previous works (e.g., \citealt{FAB2015a,Herenz2020}).
When doing this, we imagine the \lya extended emission as the result of shock fronts propagating at velocity $v_{\rm s}$ in a medium with preshock density $n_{\rm H}$. In such scenario, the \lya flux follows $F_{\rm Ly\alpha}\propto n_{\rm H}v_{\rm s}^{3}$ \citep[][]{Allen2008}. We limit the models by \citet[][]{Allen2008} to a realistic set of parameters for the nebulae here studied: $n_{\rm H}=0.01,0.1,1.0,10, 100$\,cm$^{-3}$, and $v_{\rm s}$ in the range $100 - 1000$\,km\,s$^{-1}$ in steps of 25\,km\,s$^{-1}$. We use models with solar metallicity and adopt a magnetic parameter $B/n^{1/2}=3.23$\,$\mu$G\,cm$^{3/2}$, which is expected for ISM gas assuming equipartition of magnetic and thermal energy. However, we stress that the selected models do not vary strongly with either of the two latter parameters because of the strong dependence of the ionizing flux on the shock velocity.

In more detail, in the shock models the emission is produced within the region ionized and excited by the shock.
While in the shock+precursor scenario, it is taken into account also the component ionized by the extreme ultraviolet and X-ray emission emitted upstream from the shocked region\footnote{More details on the physical mechanism at play in each model and on the production of different lines can be found in \citet[][]{Allen2008}.}.
We caution upfront that these models could underestimate the extended \lya emission since the contribution of resonant scattering is not considered. This effect would therefore shift the grids to lower values for both ratios. 

Figure~\ref{fig:line_ratios} shows the \ion{C}{iv}/Ly$\alpha$ versus the \ion{He}{ii}/Ly$\alpha$ upper limits for our sources in comparison to literature data and the model grids taken from \citet[][]{Allen2008}. The quasars studied here are shown with star symbols. The values are computed from the integrated luminosities within the nebula area (see Table~\ref{table:SFRv2}). 
The left panel shows the shock model grids while the right panel shows the shock+precursor model grids. 
We also compare our data-points to literature data for quasars' nebulae. Specifically, we indicate the line ratios for $z\sim3.5$ QSO nebulae from \citet{Borisova2016}, stacking results for nebulae around $z=3-4.5$ QSOs inside a radial bin of $10<R/{\rm kpc} < 30$ from \citet{Fossati2021}, and one extremely red quasar at $z\sim2.3$ (\citealt{Lau2022}).  
We see that for the systems studied here the shock only scenario allows for larger shock velocities ($v_{\rm s}\gtrsim300$) than the shock+precursor scenario, which restricts velocities to $v_{\rm s}\sim300-500\,{\rm km\,s^{-1}}$. 
Most of the upper limits in the literature and our most stringent data would require pre-shock densities $n_{\rm H}\gtrsim0.5\,{ \rm cm^{-3} }$ for both models. 
In the next section we show that similarly dense 
gas is needed in a photoionization scenario where \lya scattering from the quasar is neglected \citep[e.g.,][]{FAB2015b}.

\subsubsection{Photoionization from AGN followed by recombination}\label{subsec:photoionizationpower}

Unlike the quasars studied here, the SMGs do not present \lya emission at the current depth, 
indicating that AGN are vital for the powering of the observed extended nebulae. We note that even though all the QSOs in our sample are hosted by SMGs, they have a spectra consistent with the average unobscured quasar templates (e.g., \citealt[][]{Lusso2015}, see Appendix~\ref{sec:appendix_SED}). In section~\ref{subsec:scatteringpower} we discussed the likely contribution of resonant scattering of the AGN Ly$\alpha$ photons in powering the large-scale Ly$\alpha$ emission. Here, instead, we assess the impact of AGN photoionization on the observed \lya surface brightness following the framework presented in \citet[][]{Hennawi2013}. 
In this scenario, the \lya signal is produced in the recombination cascade following the gas photoionization due to the central AGN. 
The cool ($T\sim10^4$\,K) gas is assumed to be organized in small clouds with a single constant hydrogen volume density $n_{\rm H}$, constant hydrogen column density $N_{\rm H}$, and 
uniformly distributed within a spherical halo of radius $R$, such that the clouds covering factor is $f_{\rm C}$. The \lya surface brightness can then be estimated using simple relations for two limiting cases, optically thick ($N_{\rm HI}\gg10^{17.2}\,{\rm cm}^{-2}$) and optically thin ($N_{\rm HI}\ll10^{17.5}\,{\rm cm}^{-2}$) gas to ionizing radiation.

In the optically thick scenario, a thin surface of the gas clouds will emit \lya photons proportionally to the number of impinging ionizing photons (determined by the specific luminosity at the Lyman edge, $L_{\nu_{\rm LL}}$) and decreasing with distance from the quasar as $R^{-2}$. The \lya SB in the optically thick regime (${\rm SB}_{\rm Ly\alpha}^{\rm thick}$) is given by equation 15 of \citet[][]{Hennawi2013}. 

As done in \citet[][]{FAB2019b}, we estimate $L_{\nu_{\rm LL}}$ of the quasars assuming a spectral energy distribution (SED) described by a power law of the form $L_\nu=L_{\nu_{\rm LL}}(\nu/\nu_{\rm LL})^{\alpha_{\rm UV}}$ and slope $\alpha_{\rm UV}=-1.7$ as obtained by \citet[][]{Lusso2015}. 
We show a comparison of the QSOs spectra and assumed SEDs in Figure~\ref{fig:appendix_SED} from Appendix~\ref{sec:appendix_SED}, and confirm that these standard QSO templates adjust to our dataset. In particular, we stress that the two SMGs hosting a QSO in our sample are not obscured in their rest-frame UV along the line of sight, as discovered by \citet[][]{Fu2017}. We find $\log(L_{\nu_{\rm LL}}/[{\rm erg\, s^{-1}\,Hz^{-1}}])$ values of $31.8$, $30.4$ and $30.6$ for BR1202-0725, G09-0902+0101 and G15-1444-0044 respectively. We estimate the observed ${\rm SB}_{\rm Ly\alpha}^{\rm thick}$ at the same distance $R=13$\,kpc for all of the nebulae and compare to the observed values at the same distance, which is the maximum projected distance in common to all studied nebulae.  
We assume a covering factor of $f_{\rm C}=1$ for the clouds ($f_{\rm C}>0.5$ as favored by the observed diffuse morphologies, e.g., \citealt{FAB2015a}) and hydrogen column density of $N_{\rm H}=10^{20.5}\,{\rm cm^{-2}}$, as the median value obtained in absorption studies of $z\sim2$ quasar halos (\citealt{Lau2016}). In principle the covering factor could be even $f_{\rm C}>1$, indicating a clumpy medium with a larger volume filling factor. However we chose a value of 1 to simplify equations and ease comparison with previous works.
We derive ${\rm SB}_{\rm Ly\alpha}^{\rm thick}$ values of $16.5$, $2.6$ and $2.8\times10^{-15}\,{\rm erg\,s^{-1}\,cm^{-2}\,arcsec^{-2}}$ for BR1202-0725, G09-0902+0101 and G15-1444-0044 respectively. These values are a factor of $500-600$ times greater than the observed values at the same distance, which are $24.6,\ 4.3$ and $5.5{\rm \times10^{-18}\,erg\,s^{-1}\,cm^{-2}\,arcsec^{-2}}$ for BR1202-0725, G09-0902+0101 and G15-1444-0044 respectively.
Therefore we conclude that a pure optically thick regime is unlikely to be found in these systems, unless very small covering factors ($f_{\rm C}\sim0.002$) or a very small escape of Ly$\alpha$ photons are considered. This is consistent with several previous works (e.g., \citealt{FAB2015b,FAB2019b,Farina2019,Drake2020}).

Consequently, if the quasar directly shines on the surrounding gas, it is more likely in place an optically thin scenario, where the gas is highly ionized and the \lya recombination signal depends on the gas properties ($n_{\rm H}$, $N_{\rm H}$; equation 10 of  \citealt{Hennawi2013}). 
Considering the \lya SB in the optically thin regime (${\rm SB_{Ly\alpha}^{thin}}$) as the observed average SB at $R=13\,$kpc, we estimate the gas volume density $n_{\rm H}$ needed to power the observed \lya nebulae, assuming once again a covering factor of $f_{\rm C}=1$ and hydrogen column density $N_{\rm H}=10^{20.5}\,{\rm cm^{-2}}$. We find values for $n_{\rm H}$ of $13.3$, $0.6$ and $1.0\,{\rm cm^{-3}}$ for BR1202-0725, G09-0902+0101 and G15-1444-0044, respectively\footnote{We compute the density values at a specific distance neglecting dust at that location. We stress that changes in the \lya surface brightness morphology in a photoionization scenario can be due to changes in the density and dust content.}. These large ISM-like values are usually inferred by other observational studies targeting quasar nebulae \citep[see e.g.,][]{Hennawi2015,FAB2015b, Cai2018, FAB2019b}, however those studies did not have information on the dust content of the QSO's host galaxies. The density values are high for CGM gas compared to those found from simulations for $z\sim3$ massive systems \citep[][]{Rosdahl&Blaizot2012}, which are not able to resolve this gas phase.

Recent observational works showed that the high-z CGM has a multiphase nature, with even molecular gas detections on tens of kpc around some active objects
(e.g., high-redshift radio galaxies, \citealt{Emonts2016}; quasars and SMGs, \citealt{Vidal-Garcia2021}). It seems therefore realistic that the simulations need large clumping factors\footnote{The clumping factor is usually defined as $C=\langle n_{\rm H}^2\rangle/\langle n_{\rm H}\rangle^2$ \citep[e.g.,][]{Cantalupo2014}.} to match the observed \lya surface brightness on CGM scales \citep[e.g,][]{Cantalupo2014,FAB2015b,Cai2018}. However, it is still a matter of debate how strong the clumping needs to be. Large clumping factors are required when assuming only recombination ($C\sim1000$, \citealt{Cantalupo2014}), while a contribution from other mechanisms (i.e., collisional excitation and/or resonant scattering of QSO's \lya photons) would result in a more moderate clumping. 

If the photoionization scenario is the dominant mechanism, we would expect to detect, in deep observations, additional emission lines (e.g., \ion{He}{ii}) from the quasar's  CGM.
In particular, in a recombination-only scenario, the \ion{He}{ii}/\lya ratio should be $\sim 0.3$  if Helium is fully doubly ionized \citep[][]{FAB2015b,Cantalupo2019}. Lower fractions of fully ionized Helium and the contribution of additional mechanisms in the budget of \lya photons 
would result in lower ratios. Stacking analysis of $z\sim3-4$ quasars clearly detected extended \lya and \ion{C}{iv}, and only marginally \ion{He}{ii} (e.g., \citealt{Fossati2021}). That work emphasized that Cloudy photoionization models of gas on CGM scales require scattering of the quasars' \lya photons to match the observed line ratios using reasonable values for CGM metallicities. Since we detect \lya emission and do not detect \ion{He}{ii}, resulting in similar upper limits as \citet{Fossati2021} (\ion{He}{ii}/\lya~$<0.03$~), we cannot provide further insights on this problem. Future works are needed to establish the relative contribution of recombination and QSO's \lya scattered photons to the \lya glow on CGM scales. 

\subsection{Comparison with previous Ly$\alpha$ SB radial profiles}\label{subsec:SBprofilesDiscussion}

\begin{figure*}
    \centering
    \includegraphics[width=1.0\textwidth]{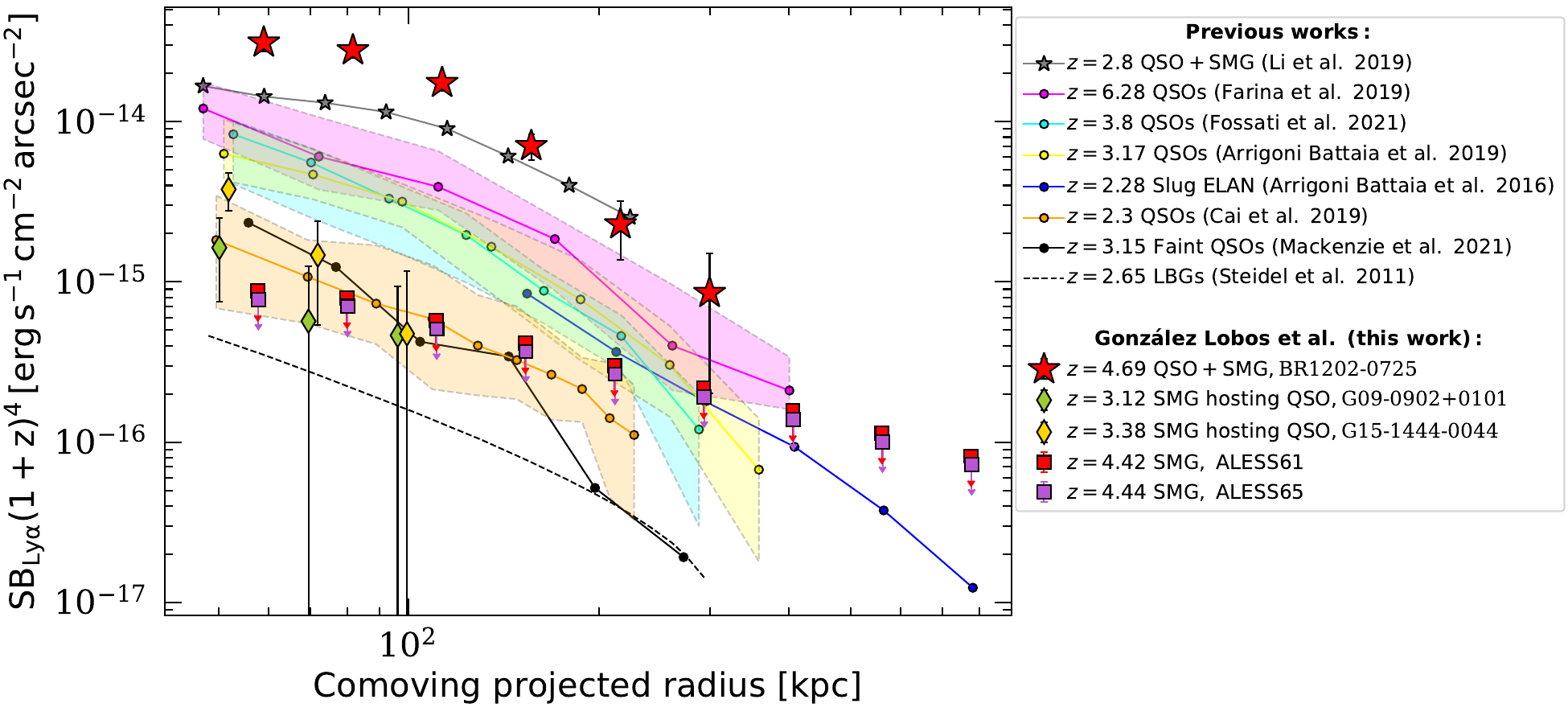}
    \caption{Circularly averaged SB as a function of comoving projected radius of the nebulae detected in our quasar sample (red stars, green and yellow diamonds for $\rm{BR1202\textsc{-}0725}$, $\rm{G09\textsc{-}0902\textsc{+}0101}$ and $\rm{G15\textsc{-}1444\textsc{-}0044}$, respectively), corrected for cosmological dimming. The red and purple squares with arrows indicate the upper limits 
    for ALESS61.1 and ALESS65.1, respectively. Different profiles from the literature are indicated as colored circles with their respective $25-75$ percentile ranges when available. The ${z=3.17}$ sample of the QSO MUSEUM from \citet{FAB2019} is shown in yellow. 
    The SB profile (taken from \citealt{FAB2016}) of the $z=2.28$ Slug Enormous \lya Nebula (ELAN) is shown in blue (\citealt{Cantalupo2014}). The ${z=6.2}$ quasar sample from \citet{Farina2019} is shown in 
    magenta. 
    The ${z=3.8}$ quasar sample from \citet{Fossati2021} is shown in 
    cyan. 
    The ${z=2.3}$ quasar sample from \citet{Cai2019} is shown in 
    orange. 
    The median profile of the faint quasar sample at $z=3$ from \citet[][]{Mackenzie2020} is shown with black connected circles. We computed the SB profile of the galaxy group SMMJ02399 comprising a quasar and an SMG at ${z=2.8}$ from \citet[][]{Li2019} (gray connected circles). Finally, the fit to the observed SB profile for $z=2.65$ LBGs from \citet[][]{Steidel2011} is shown with the black dashed curve.}
    \label{fig:radialprofiles}
\end{figure*}

Figure~\ref{fig:radialprofiles} shows the SB radial profiles of the three detected targets as red stars, green and yellow diamonds for $\rm{BR1202\textsc{-}0725}$, $\rm{G09\textsc{-}0902\textsc{+}0101}$ and $\rm{G15\textsc{-}1444\textsc{-}0044}$, respectively, as a function of comoving projected radius, and corrected by cosmological dimming. The profiles are shown only for annulus above the ${\rm 2\sigma}$ SB limit for that ring (red line of Figure~\ref{fig:maprofiles}). Once again, it is evident that the nebulae associated with the SMGs hosting a QSO 
are dimmer and more compact than the one around the QSO+SMG system. Further, the same figure shows a comparison with different samples from the literature (see figure caption and legend), highlighting the respective ${\rm 25-75}$ percentile ranges when available. We find that the sources in our sample are outliers for any of the ${z\gtrsim3}$ bright quasar samples \citep[][]{FAB2019, Farina2019, Fossati2021}. However, we find that our faint, but submillimeter-bright ${z\sim3}$ quasars present nebulae akin to those found around similarly faint ($i_{\rm mag}<19.5$) ${z\sim3}$ quasars (\citealt{Mackenzie2020}), and that resemble the median profile from the lower redshift (${z\sim2.3}$) sample from \citet{Cai2019}. This once again stresses the large difference between nebulae of bright ${z\sim2.3}$ quasars and those of bright ${z\sim3}$ quasars, which has been argued to be most likely driven by a different halo accretion regime (e.g., \citealt{FAB2019,Farina2019,Fossati2021}). For completeness, we also show the observed average SB profile of $z=2.65$ Lyman Break Galaxies (LBGs) from \citet[][]{Steidel2011}. These galaxies are sources at the lower end of the halo masses probed by our sample (median halo mass of $M_{\rm DM} = 9\times10^{11}\,{\rm M_\odot}$; \citealt{Steidel2011}), and show both signatures of \lya emission and absorption on galaxy scales, but they do not have an AGN contribution. LBGs clearly have an SB profile fainter than sources hosting QSOs, further stressing the importance of AGN in powering strong \lya emission.

On the other hand, the QSO+SMG system ${\rm BR1202\textsc{-}0725}$ has clearly an enhanced brightness with respect to all other profiles. This could be due to several factors, like (i) the merger state of this system with interacting galaxies (an SMG, a QSO host and a LAE), likely providing a larger reservoir of cool gas with respect to individual halos and possible supply for diffuse star formation (e.g., \citealt{Decarli2019}), (ii) the presence of dense gas on CGM scales, as directly evident from the presence of  
a bridge of [\ion{C}{ii}] emission between the QSO and the SMG \citep[][]{Drake2020}, and (iii) presence of galactic and AGN winds/outflows that could enhance densities in their surroundings and/or favor the escape of Ly$\alpha$ photons (e.g., \citealt{Costa2022}).
To confirm whether brighter SB profiles are typical for QSO+SMG systems, we extract the radial profile for the SMMJ02399 system at ${z=2.8}$ (gray points) from \citet{Li2019}. This system is also known to have a multiphase CGM with a cold diffuse molecular phase extending out to at least $\sim 20$~kpc (\citealt{Vidal-Garcia2021}). We also find that this system is at the high end of the radial profiles, even though its redshift is lower.
This finding further suggests that QSO+SMG systems are the pinnacle of \lya emission at each redshift. Further evidence of this comes from the comparison of the SB radial profile of the $z\sim2.28$ Slug Enormous \lya Nebula (ELAN, blue points) from \citet{FAB2016} with the profiles of similar bright quasars at $z\sim2$ \citep[][]{Cai2019}. The Slug ELAN is known to host an SMG undergoing ram-pressure stripping within the halo of a bright QSO (\citealt{Chen2021}). Similar to the aforementioned cases, the SB profile of the Slug ELAN is above the average $z\sim2$ quasars' profile.
However, its SB profile is not as high as the one found for ${\rm BR1202\textsc{-}0725}$ and SMMJ02399, possibly indicating a difference in the densities and reservoir of cool material around these systems from high to low redshift (e.g., \citealt[][]{FAB2019}). 
In the same figure we also indicate the ${\rm 2\sigma}$ upper limits for the SB inside each ring for the SMGs, as the red (${\rm ALESS61.1}$) and purple (${\rm ALESS65.1}$) squares with arrows. We find that the \lya SB for isolated SMGs is at least a factor of 2 fainter than that of quasars at similar redshifts.

Finally, we compare the scale length $r_{\rm h}=11\pm1$~kpc obtained for BR1202-0725 in Section~\ref{subsec:radialprofiles} with those reported for the average profiles of type-I quasars at bracketing redshifts $z\sim3$ and $z\sim6$. We find that BR1202-0725 sitting at $z=4.6942$ has a scale length smaller (larger) than the sample at $z\sim3$ ($z\sim6$) when looking at physical scales, $r_{\rm h}=15.7$~kpc (\citealt{FAB2019}) and $r_{\rm h}=9.4$ (\citealt{Farina2019}), respectively. This would translate to similar scale lengths in comoving units as can be appreciated in Figure~\ref{fig:radialprofiles}. Larger samples of quasar's nebulae at $z\sim5$ are needed to firmly constrain these values.

%%%%%%%%%%%%%%%
%   SUMMARY   %
%%%%%%%%%%%%%%%

\section{Summary}\label{sec:summary}

With VLT/MUSE, we targeted five ${\rm z\sim3-4}$ submillimeter-bright systems to unveil extended \lya emission, discuss the contribution of different powering mechanisms, 
as well as compare our results to the plethora of ${\rm z>2}$ quasar observations. 
Specifically, our observations comprise systems with different degrees of quasar illumination: a QSO+SMG system (BR1202-0725), two SMGs hosting a QSO (G09-0902+0101 and G15-1444-0044), and two SMGs (ALESS61.1 and ALESS65.1). All these objects are 
expected to sit in halos as massive as those hosting quasars (section~\ref{sec:data}).
Below, we summarize our observational findings:

\begin{itemize}
    \item We find that our targets have very different \lya morphologies and brightness levels. The QSO+SMG system presents the highest levels of \lya emission, probably reflecting the richness of its halo environment in terms of mass and density of the cool gas phase (top panel of Figure~\ref{fig:sbmaps-qso} and section~\ref{subsec:SBprofilesDiscussion}). 
    We do not detect any extended \lya emission around the two SMGs.
    \item All the detected \lya nebulosities present more violent kinematics (FWHM$\gtrsim1200$\,km\,s$^{-1}$) with respect to nebulae around quasars at similar redshifts (FWHM$\sim600$\,km\,s$^{-1}$; \citealt[][]{FAB2019,Fossati2021}). In particular, we find evidence of large-scale outflows in the BR1202-0725 system (Figure~\ref{sec:BR1202_kinematics}).
    \item None of the targeted systems show extended emission in other rest-frame UV lines besides \lya down to $\sim 10^{-18}$~erg~s$^{-1}$~cm$^{-2}$~arcsec$^{-2}$ ($2\sigma$ in 30~\AA; Section~\ref{sec:otherLines}).
    \item The two SMGs hosting a QSO have CGM \lya profiles about 3 times fainter than the typical ${\rm z\sim3}$ quasar halo, resembling the median profile of similarly faint quasars \citet[][]{Mackenzie2020} (Figure~\ref{fig:radialprofiles}; section~\ref{subsec:SBprofilesDiscussion}). We also find that their \lya emission is at similar levels than the emission around bright ${\rm z\sim2}$ quasars, further indicating the remarkable difference between the \lya emission around bright ${\rm z\sim2}$ quasars and higher redshift quasars (e.g., \citealt[][]{FAB2019,Farina2019,Fossati2021}). 
\end{itemize}

We then discuss our observational results in the context of the relative roles of the different \lya powering mechanisms presented in Section~\ref{sec:intro}: resonant scattering of \lya photons for embedded sources, gravitational cooling radiation, shocks by galactic/AGN outflows, and photoionization from AGN or star-formation followed by recombination. We briefly summarize our results here.

Cooling radiation is unlikely to be the main powering mechanism, because  
we would expect the \lya emission to scale proportionally to the halo mass (see Section~\ref{subsec:coolingpower}). Therefore, we should have detected similar levels of \lya emission for the similar halo masses of our sample \citep[e.g.,][]{Haiman2000,Rosdahl&Blaizot2012}. We propose that more observations of obscured systems, such as isolated SMGs, could provide test cases for this scenario.

We cannot rule out that resonant scattering of \lya photons emitted by compact sources could have a significant contribution to the observed nebulae. 
Indeed, we observe that the extended \lya level is consistent with being regulated by the budget of QSO \lya photons modulo the dust content on galaxy scales. Sources with no \lya emission on small scales, i.e. the SMGs, do not show extended \lya emission. However, due to the small size of our sample we cannot provide firm constraints on the relationship between the extended \lya luminosity and the budget of QSO \lya photons. Therefore larger samples of QSO \lya nebulae with information on their host galaxies are needed.

Regarding the star-formation scenario, by comparing the inferred ${\rm SFR_{Ly\alpha}}$ between our nebulae (Table~\ref{table:SFRv2}) to the derived SFR from their total IR luminosity (Table~\ref{sec:data}), we show that the available star-formation in these systems should be enough to power the \lya nebulae (see Section~\ref{subsec:starformpower}). However, the fact that we do not detect extended \lya emission around the SMGs suggests that photoionization due to star-formation followed by recombination (and resonant scattering) has a minor role in powering these nebulae likely because most of the UV and \lya photons from star-formation do not escape these dusty galaxies.

Also, the aforementioned presence of large \lya linewidths and positive velocity shifts (Table~\ref{table:SFRv2}) of the QSO's nebulae motivates us to test a large-scale outflow scenario in Section~\ref{subsec:shockpower}. We compare the \ion{C}{iv}/\lya and \ion{He}{ii}/\lya line-ratio upper limits with shock and shock+precursor models from \citet[][]{Allen2008}. We find that the ratios for our systems agree with the models with shock velocities $v_{\rm s} \geq 300\,{ \rm km\,s^{-1} }$ for the shock only scenario and $v_{\rm s} \sim 300-500\,{ \rm km\,s^{-1} }$ for the shock+precursor scenario. Both cases allow for a high pre-shock density of $n_{\rm H}\geq1\,{\rm cm^{-3}}$, usually not resolved in cosmological simulations and expected within the interstellar-medium.

Moreover, we find that the AGN presence is essential for powering the observed \lya halos, as only systems with quasars show an extended \lya glow, highlighting the importance of photoionization from the central AGN as a contribution to the extended \lya emission. We test the contribution of photoionization from the quasar followed by recombination in Section~\ref{subsec:photoionizationpower}, under the assumption of two limiting scenarios: optically thick and optically thin to ionizing radiation. Assuming the quasar is illuminating the gas, we find that our data favors an optically thin scenario, which requires an average volume density of $n_{\rm H}\sim1-10\,{ \rm cm^{-3} }$, as frequently found in previous studies (e.g., \citealt[][]{FAB2015a,Hennawi2015,Cai2018}).  

Summarizing, we find evidence against gravitational cooling radiation to be the main powering mechanism of the detected extended \lya emission. On the other hand, photoionization, outflows and resonant scattering of \lya photons from compact sources are likely contributors to the observed \lya nebulae.
Our pilot work opens the path to the analysis of CGM emission around high-z massive systems taking into account the properties of quasar's hosts (e.g., \citealt{Nahir2022}) and embedded galaxies, frequently overlooked in the study of large-scale emission. Future large statistical samples together with statistical investigations of mock observations are needed to confirm and extend our findings. Moreover, future JWST observations of extended H$\alpha$ emission around QSOs could help disentangling the powering mechanisms discussed in this work.

\begin{acknowledgements}

The authors thank the anonymous referee for providing constructive comments that have improved our paper.
The authors thank Kevin Hall for his useful comments on an advanced draft of this work.
Based on observations collected at the European Organisation for Astronomical Research in the Southern Hemisphere under ESO programmes 0103.A-0296(A), 0102.A-0403(A) and 0102.A-0428(A), available from the ESO Science Archive Facility \url{https://archive.eso.org/}.
EPF is supported by the international Gemini Observatory, a program of NSF's NOIRLab, which is managed by the Association of Universities for Research in Astronomy (AURA) under a cooperative agreement with the National Science Foundation, on behalf of the Gemini partnership of Argentina, Brazil, Canada, Chile, the Republic of Korea, and the United States of America. C.-C.C. acknowledges support from the National Science and Technology Council of Taiwan (NSTC 109-2112-M-001-016-MY3 and 111-2112-M-001-045-MY3), as well as Academia Sinica through the Career Development Award (AS-CDA-112-M02). A.O. is funded by the Deutsche Forschungsgemeinschaft (DFG, German Research Foundation) – 443044596. M.G. thanks the Max Planck Society for support through the Max Planck Research Group.

\end{acknowledgements}

% ------------------------------

%\section*{Data Availability}

%The observational data used in this work are available from the ESO Science Archive Facility \url{https://archive.eso.org/}.

% ------------------------------

\bibliographystyle{aa}
\bibliography{example}

\begin{appendix}

\section{Seeing as a function of wavelength}\label{sec:appendix-seeing}

As mentioned in Section~\ref{sec:analysis}, we subtract the wavelength-dependent unresolved quasar's PSF in order to reveal extended \lya emission. The PSF is subtracted centered at the quasar position and out to six times the seeing, this value is chosen so that the PSF profile becomes consistent with zero. To estimate the seeing we perform a 2D Moffat fit to pseudo narrow-bands (width of 25\,\AA) of stacked point sources in each field as a function of wavelength.

The 2D Moffat model is given by:
\begin{equation}
    f(x,y) = A\left(1+\dfrac{(x-x_0)^2+(y-y_0)^2}{\gamma^2}\right)^{-\beta},
\end{equation}
where $A$ is the scale amplitude, $x_0$ and $y_0$ are the point source centroid coordinates, $\gamma$ is the core width of the model, and $\beta$ is the power of the model. The seeing is obtained from the FWHM of the Moffat profile, given by ${ \rm FWHM=2 \gamma \sqrt{ 2^{1/\beta} - 1 } }$.
We show in Table~\ref{tab:Data} the seeing at the expected \lya wavelength of each field.

\begin{figure}
    \centering
    \includegraphics[width=0.45\textwidth]{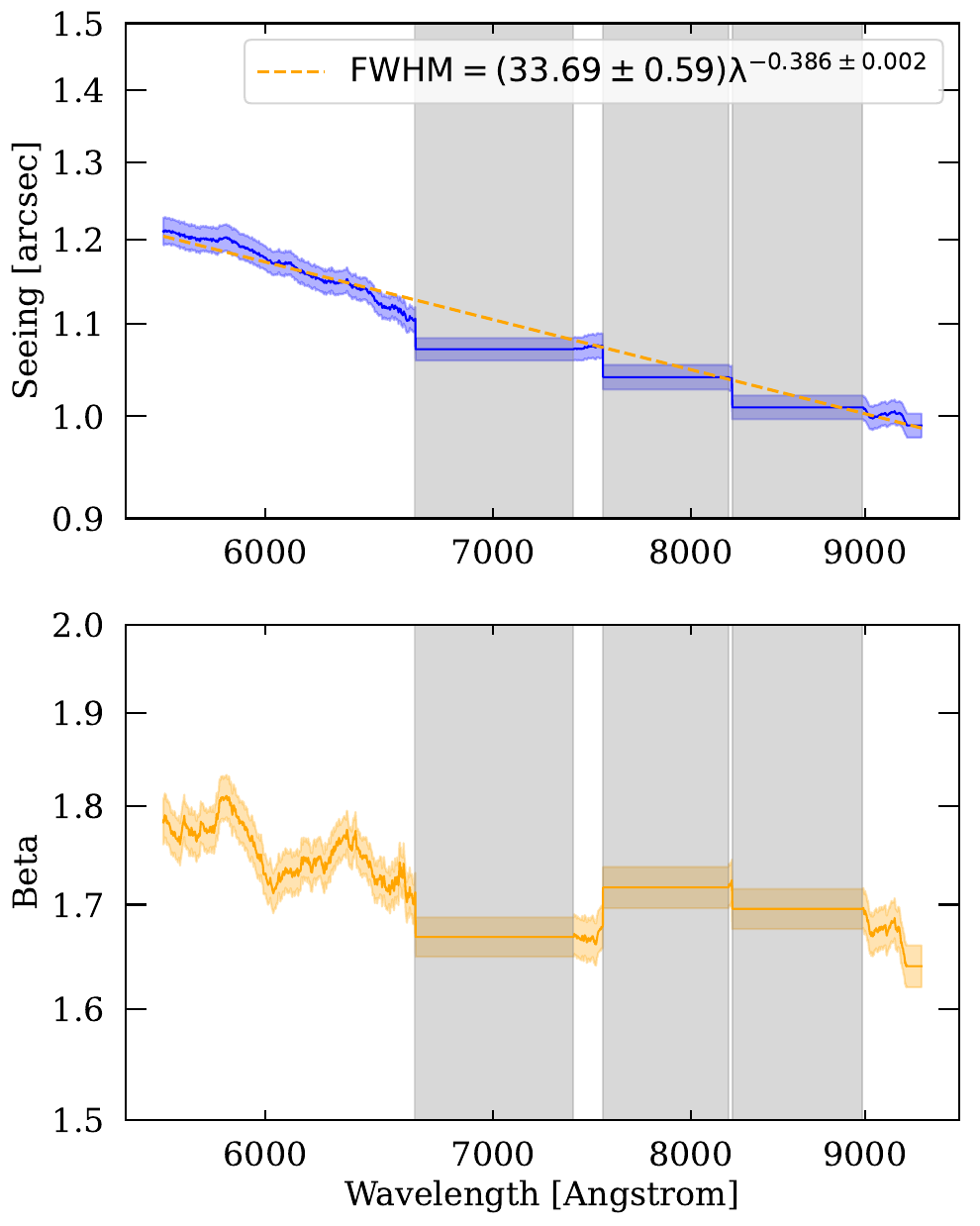}
    \caption{Properties of the wavelength dependent Moffat fit used to estimate the seeing for BR1202-0725. {\it Top:} Seeing (FWHM of the Moffat fit) as a function of observed wavelength of BR1202-0725's PSF (blue) with its respective $1\sigma$ uncertainty as shaded region. The power law fit is shown with a dashed orange line and the fitted parameters are shown in the legend. {\it Bottom:} Beta parameter of the fitted Moffat profile as a function of wavelength with its respective $1\sigma$ uncertainty.}
    \label{fig:seeing}
\end{figure}

We show in Figure~\ref{fig:seeing} an example of the wavelength-dependent FWHM and $\beta$ of one of the quasar's empirical PSF, i.e. for BR1202-0725.
We exclude wavelengths significantly affected by the Ly$\alpha$ forest ($<5600$\, \AA) and indicate with gray shaded regions the wavelength ranges where extended Ly$\alpha$ emission is conservatively expected and therefore the PSF is kept constant (see Section~\ref{subsec:psfsub} and Appendix~\ref{sec:lines_identification}).
We find an expected decrease with wavelength of the FWHM and a close to constant $\beta$ parameter consistent with the trend and values found in previous MUSE studies \citep[$\beta\sim1.7$; e.g.,][]{Fusco2020}. In Figure~\ref{fig:seeing} we also show a power-law fit to the FWHM as a function of wavelength, which is consistent with the seeing at the \lya wavelength reported in Section~\ref{tab:Data} from point sources.
This fit excludes the gray shaded regions as these are constant by construction.

\section{Line emission identification}\label{sec:lines_identification}

\begin{figure}
    \centering
    \includegraphics[width=1.07\columnwidth]{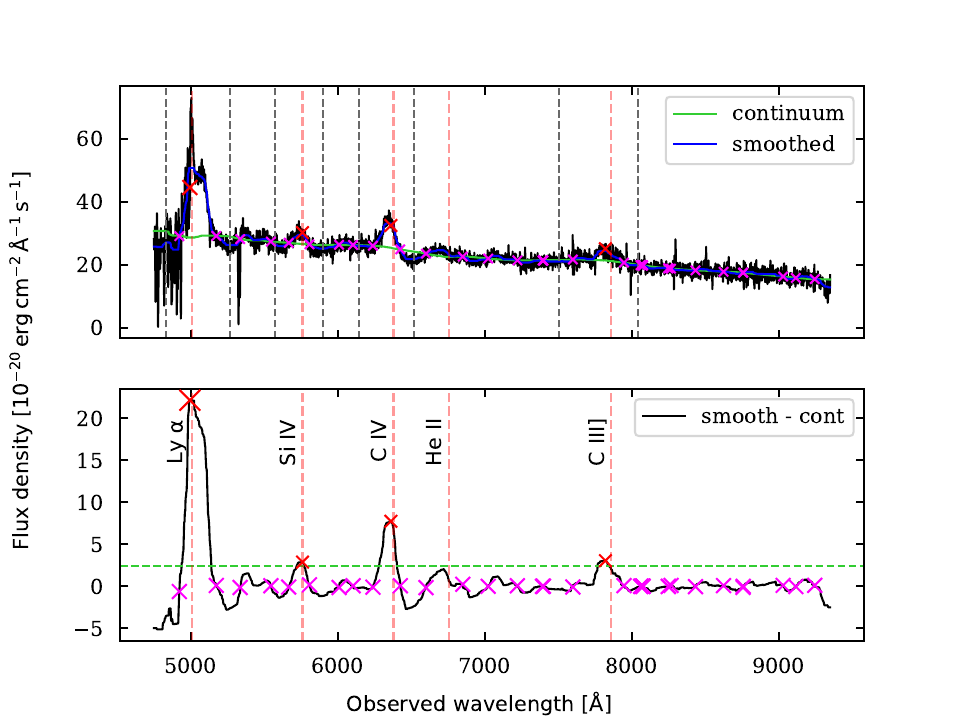}
    \caption{Example of the quasar's broad line emission identification for G09-0902+0101. \textit{Top:} Integrated spectrum inside a 1 arcsecond diameter aperture of G09-0902+0101 is shown in black. Overlayed are a smoothed spectrum (blue) and "continuum" spectrum (green), constructed by using a median filter of 50 and 500 channels respectively. The red and pink crosses indicate detected peaks of line emission and the intersection between the smooth and "continuum" spectra respectively. The vertical gray lines indicate the regions where each line emission is excluded to build the PSF (see Section~\ref{subsec:psfsub}). \textit{Bottom:} Difference spectrum between the smooth and continuum spectrum of the top panel. The green dashed line indicates the $3\sigma$ threshold for line identification. We indicate with a red dashed line the position of the quasar's \lya and broad line emissions, with their respective label.}
    \label{fig:lines_iden}
\end{figure}

As mentioned in Section~\ref{subsec:psfsub}, in order to not subtract the extended emission we are interested in, we construct the wavelength-dependent empirical PSF of each quasar only where there is no line emission. For this, we create a routine that identifies peaks of line emission at the expected wavelengths. This routine is only intended to identify the location in wavelength of the line emissions, instead of characterising the line emission of the quasar. We detect line emission by integrating a spectrum at the quasar location inside a 1 arcsecond diameter aperture, shown in Figure~\ref{fig:lines_iden}. We then use a median filter to create two smoothed spectra, one that highlights the QSO spectrum broad lines (smooth spectrum) and another that smooths out the broad line features ("continuum" spectrum\footnote{In practice, this is not intended to be a perfect characterization of the quasar continuum, but it allows us to quickly identify the location of the line emission.}). The smooth spectrum uses a median filter of 50 spectral channels (62.5\,\AA), while the "continuum" spectrum uses a median filter of 500 spectral channels (625\,\AA). We find where these two spectra intersect by creating a difference spectrum between the smooth and continuum spectrum, we then identify the lines by finding positive peaks in the difference spectrum that are above a $3\sigma$ threshold. The peaks are found using \texttt{scipy.signal.find$\_$peaks} in combination with the $3\sigma$ threshold and a minimum separation between the local peaks. The $3\sigma$ threshold is computed from the RMS of the difference spectrum redwards of the \ion{C}{iv} line, because we do not expect to detect more line emissions at these wavelengths due to the redshift of our sources. In Figure~\ref{fig:lines_iden} we show an example of line emission identification for G09-0902+0101. Where the top panel shows the spectrum and the locations of line emission are marked with red crosses, the vertical dashed lines show the regions where we find the wavelength ranges for each line. These ranges are excluded when we build the empirical PSF, and instead use the next available PSF redwards of the line emission.

\section{\ion{Si}{iv}, \ion{C}{iv}, \ion{He}{ii} and \ion{C}{iii]} analysis}\label{sec:appendix_lines}

\begin{figure*}
    \centering 
    \includegraphics[width=0.8\textwidth]{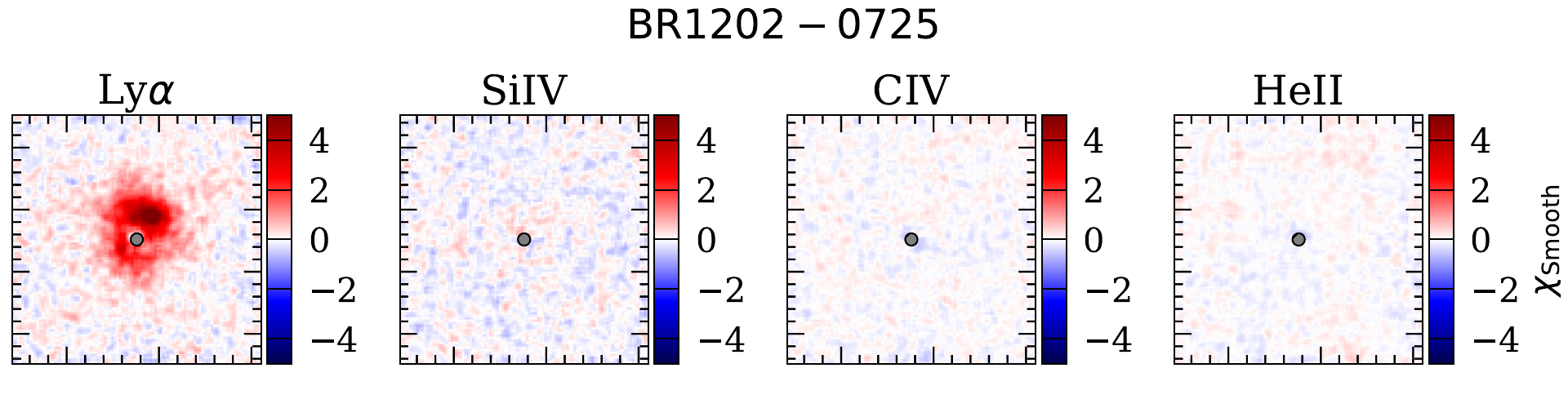}
    \caption{\lya, \ion{Si}{iv}, \ion{C}{iv} and \ion{He}{ii}  $\chi_{\rm Smooth}$ maps using a velocity window equivalent to the 30\,\AA\ narrow-band used for the \lya SB maps for BR1202-0725.
    The side of the maps is 20\arcsec, corresponding to 129\,kpc. We show with a circle the $1\arcsec\times1\arcsec$ normalisation area of the quasar's PSF-subtraction.}
    \label{fig:linesBR1202}
\end{figure*}

\begin{figure*}
    \centering 
    \includegraphics[width=\textwidth]{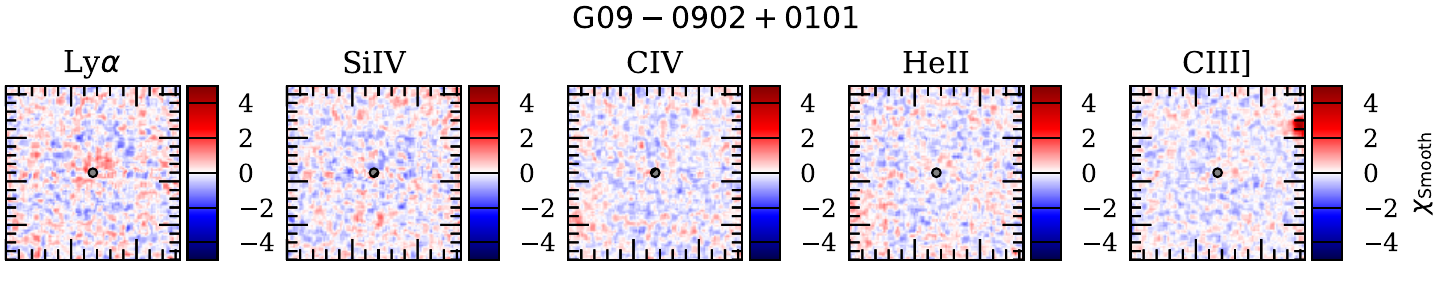}
    \caption{\lya, \ion{Si}{iv}, \ion{C}{iv}, \ion{He}{ii} and \ion{C}{iii]} $\chi_{\rm Smooth}$ maps using a velocity window equivalent to the 30\,\AA\ narrow-band used for the \lya SB maps for G09-0902-0101. The side of the maps is 20\arcsec, corresponding to 152\,kpc. We show with a circle the $1\arcsec\times1\arcsec$ normalisation area of the quasar's PSF-subtraction.}
        \label{fig:linesG09}
\end{figure*}

\begin{figure*}
    \centering 
    \includegraphics[width=\textwidth]{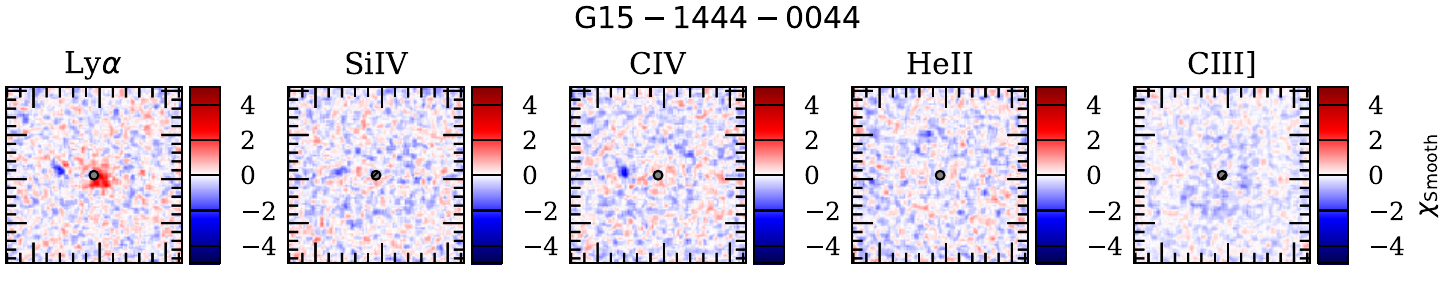}
    \caption{Same as Figure~\ref{fig:linesG09}, but for G15-1444-0044. The 20\arcsec side of the maps corresponds to 148\,kpc.} 
        \label{fig:linesG15}
\end{figure*}

\begin{figure*}
    \centering 
    \includegraphics[width=0.75\textwidth]{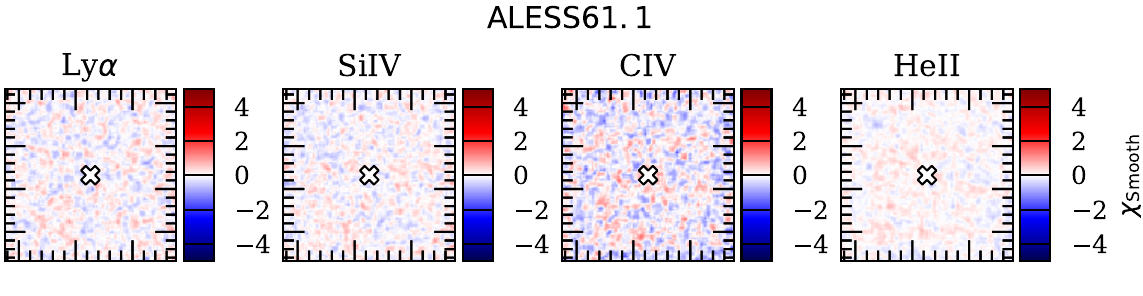}
    \caption{\lya, \ion{Si}{iv}, \ion{C}{iv}, and \ion{He}{ii} 
    $\chi_{\rm Smooth}$ maps using a velocity window equivalent to the 30\,\AA narrow-band used for the \lya SB maps for ALESS61.1. The side of the maps is 20\arcsec, corresponding to 133\,kpc. We show with a white cross the position of the SMG.}
        \label{fig:linesA61}
\end{figure*}

\begin{figure*}
    \centering 
    \includegraphics[width=0.75\textwidth]{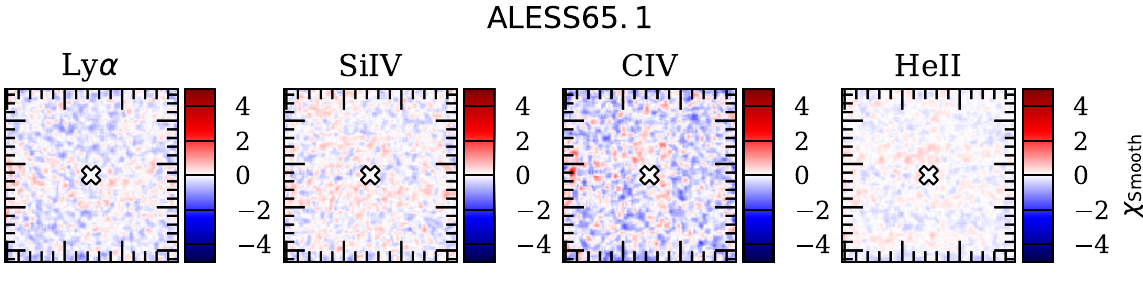}
    \caption{Same as Figure~\ref{fig:linesA61}, but for ALESS65.1. The 20\arcsec side of the maps corresponds to 132\,kpc.}
        \label{fig:linesA65}
\end{figure*}

We conduct a search for other extended line emission besides Ly$\alpha$, such as \ion{Si}{iv}, \ion{C}{iv}, \ion{He}{ii} and \ion{C}{iii]}. However, we do not find extended emission in these lines for any target (Section~\ref{sec:otherLines}). For completeness, we build $\chi_{\rm Smooth}$ maps (e.g., \citealt{Hennawi2013}) centered at the expected wavelength of each line and using a narrow-band window that reflects the 30\,\AA\ narrow-band used for \lya. The $\chi_{\rm Smooth}$ maps for the systems studied here are shown in Figures~\ref{fig:linesBR1202},~\ref{fig:linesG09},~\ref{fig:linesG15},~\ref{fig:linesA61}~and~\ref{fig:linesA65}, these maps are best suited for visualizing the presence of extended emission (if any), and to appreciate the noise in the data. 
The $\chi_{\rm Smooth}$ maps are computed using equation (2) from \citet[][]{Farina2019}, given by:

\begin{equation}
    {\rm SMOOTH}[\chi_{x,y,\lambda}] = \dfrac{ {\rm CONVOL}[{\rm DATA}_{x,y,\lambda} - {\rm MODEL}_{x,y,\lambda} ] }{ \sqrt{{\rm CONVOL^2}[\sigma^2_{x,y,\lambda}]} },
\end{equation}
where ${\rm DATA}_{x,y,\lambda}$ is the original datacube, ${\rm MODEL}_{x,y,\lambda}$ is the empirical PSF of the quasar obtained in Section~\ref{subsec:psfsub}, $\sigma^2_{x,y,\lambda}$ is the variance datacube. ${\rm CONVOL}$ is a spatial convolution using a 2D box kernel with of 3 pixels width (0.6\arcsec) and ${\rm CONVOL^2}$ is a spatial convolution using the square of the width of the kernel used for ${\rm CONVOL}$. 
Different noise properties depends also on the vicinity to sky lines. For example, the \ion{He}{ii} velocity range of BR1202-0725 displays sky lines, therefore we subtract the residuals from the data before computing the $\chi_{\rm Smooth}$ map.

\section{QSO spectral energy distribution fitting}\label{sec:appendix_SED}

\begin{figure}
    \centering
    \includegraphics[width=0.45\textwidth]{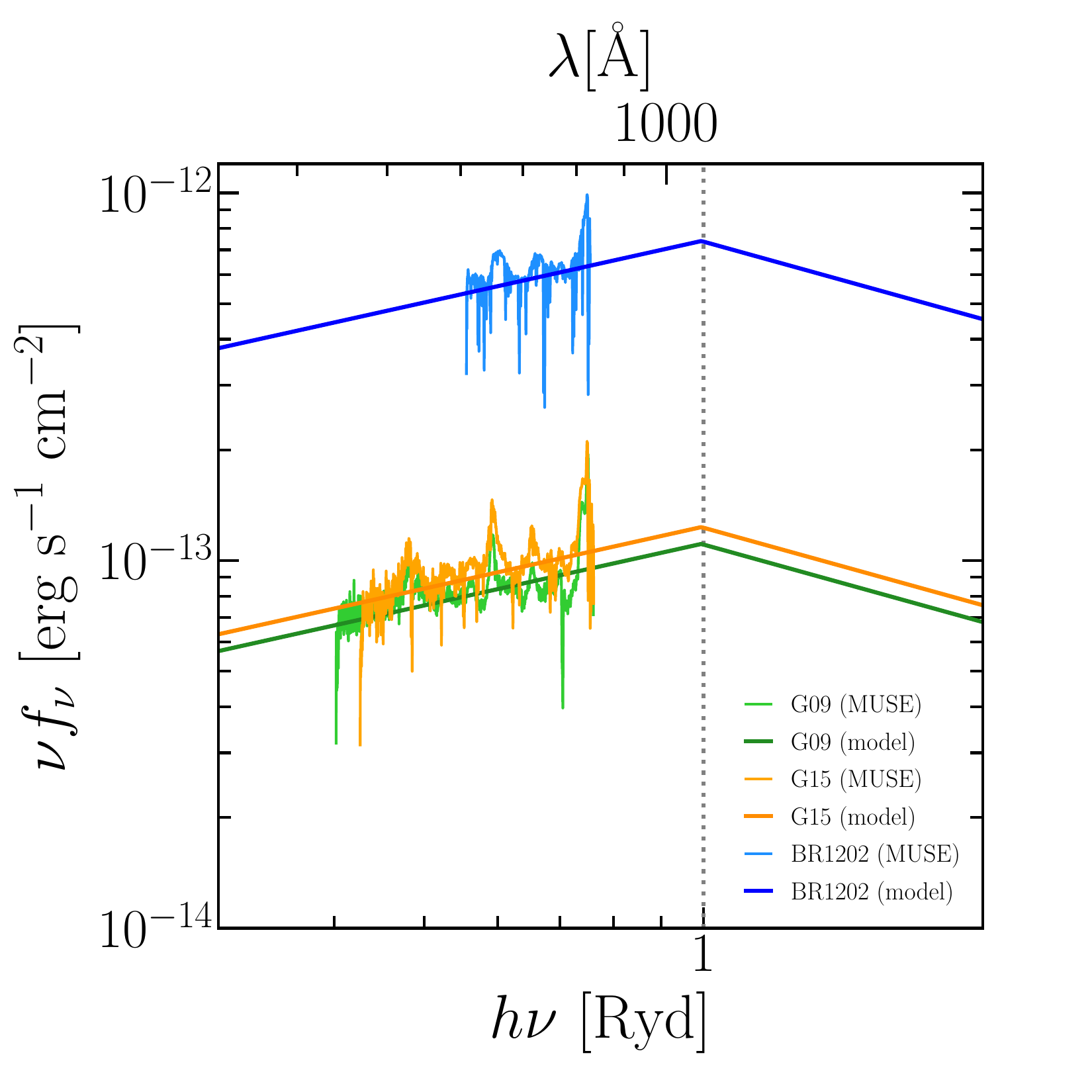}
    \caption{Comparison of the SED of the three quasars studied here used as incident radiation field in the modeling. The blue, orange and green solid lines indicate the input spectrum with slope $\alpha_{\rm UV}=-1.7$ \citep[][]{Lusso2015} for BR1202-0725, G09-0902+0101 and G15-1444-0044 respectively. The curves of the same color indicate the MUSE spectrum of each quasar extracted within a 1.5\arcsec radius aperture.}
    \label{fig:appendix_SED}
\end{figure}

As mentioned in Section~\ref{subsec:photoionizationpower}, we model the number of quasars' ionizing photons impinging on optically thick gas estimating the specific luminosity at the Lyman edge ($L_{\rm \nu_{LL}}$) for the three quasars in our sample. As shown in \citet[][]{FAB2019b}, a QSO SED can be parametrized by different power-laws, with the extreme ultra-violet portion described by a power law of the form $L_\nu=L_{\rm \nu_{LL}}(\nu/\nu_{\rm LL})^{\rm \alpha_{UV}}$ with slope $\alpha_{\rm UV}=-1.7$ (as obtained in \citet[][]{Lusso2015}), and a slope $\alpha_{\rm opt}=-0.46$ below 1\,Ryd (\citealt{vandenBerk2001}). Figure~\ref{fig:appendix_SED} shows a comparison of the MUSE QSO spectra and the assumed SED (following \citealt{FAB2019b}), which confirms that a standard QSO template adjust to the sample studied here.

\end{appendix}

\end{document}